\documentclass[
preprint,
preprintnumbers,
 amsmath,amssymb,
 aps,
showkeys]{revtex4-2}

\usepackage{graphicx}
\usepackage{dcolumn}
\usepackage{bm}
\usepackage{siunitx}
\usepackage[noabbrev,capitalize]{cleveref}
\usepackage{subcaption}
\usepackage{tikz}
\usepackage{chemfig}
\setchemfig{atom sep=1.5em}

\begin{document}


\title{Effects of Epoxy Composition on the Thermal and Network Properties of Crosslinked Thermosets: A Molecular-Dynamics Study}

\author{Manoj Pola}
\email{v.r.m.pola@tue.nl}
\affiliation{Soft Matter and Biological Physics group, Department of Applied Physics and Science Education, Eindhoven University of Technology}

\author{Alexey Lyulin}

\affiliation{Soft Matter and Biological Physics group, Department of Applied Physics and Science Education, Eindhoven University of Technology}
\affiliation{Multiscale Molecular-Dynamics Group
Faculty EEMCS,
University of Twente}

\date{\today}

\begin{abstract}
In this study, we investigate the influence of resin composition on the thermal and structural properties of crosslinked epoxy networks using molecular-dynamics simulations. The systems studied are based on mixtures of diglycidyl ether of bisphenol A (DGEBA) and butylated epoxy resin (BER), cured with diethyltoluenediamine (DETDA). A multi-step crosslinking algorithm was used to generate highly crosslinked networks. We systematically explored how the addition of BER affects the network topology and thermophysical behavior of the resulting thermosets.

Our results show that increasing BER content leads to a progressive loosening of the network, evidenced by longer inter-nitrogen path lengths and larger Voronoi atomic volumes. This structural loosening results in a higher coefficient of thermal expansion (CTE) and higher atomic mobility as well as a reduction in the glass-transition temperature ($T_g$). To quantify local steric and topological effects, we introduce a confinement index that captures atomic-level crowding and connectivity. The index shows a strong correlation with the observed trend in $T_g$, suggesting its potential as a predictive descriptor for polymer network behavior. Our findings provide molecular-level insights into how epoxy resin composition modulates network structure and thermomechanical properties.

\end{abstract}

\maketitle


\section{Introduction}

The term epoxy refers to a class of chemicals whose monomers contain one or more terminal epoxide groups. An epoxide group is a three-membered ring consisting of one oxygen atom and two carbon atoms. Literature on the synthesis of epoxies dates back to the early 1900s~\cite{lee1967handbook}. In the presence of a suitable crosslinker, epoxies undergo curing to form thermosets. These thermosets exhibit excellent mechanical, chemical, and thermal properties, making them ideal for a wide range of applications, including coatings, adhesives, encapsulants, and composite matrices~\cite{pham2000epoxy}. According to Fortune Business Insights, the global epoxy market was valued at USD 12.74 billion in 2023~\cite{fortunebusinessinsightsEpoxyResin}. The physical characteristics of epoxy systems can be tailored by modifying the resin composition, introducing additives, and selecting different crosslinkers, among other factors~\cite{may2018epoxy}.

In the present study, we simulate mixtures of two epoxy resins: diglycidyl ether of bisphenol A (DGEBA, CAS No. 1675-54-3) and butylated epoxy resin (BER, CAS No. 71033-08-4), using diethyltoluenediamine (DETDA, CAS No. 2095-02-5) as the crosslinker. Mixtures of DGEBA and BER are commonly employed in adhesive, encapsulation, and sealant applications~\cite{epotek_t7109_sds,stebbins_sds116_2023}. BER is typically added to improve the flexibility of the cured thermoset. Although specific literature on the use of BER is limited, \citet{Monte1998} discussed the incorporation of various additives into epoxy resins, mentioning chemicals similar to BER as flexibility-enhancing agents. The present study, however, does not focus on flexibility. Instead, it investigates the thermal, structural, and topological properties of the cured thermosets.

Molecular-dynamics (MD) simulations have been employed to model and analyze the crosslinked systems formed from DGEBA, BER, and DETDA monomers. Hereafter, the terms system or sample refer to the molecular topologies used in the simulations. One of the earliest uses of MD to model crosslinked polymers was by \citet{doherty1998polymerization}, who studied poly(methacrylate) networks generated using a single-step, cutoff-radius-based crosslinking scheme. A similar method was later employed by \citet{yarovsky2002computer} to investigate low-molecular-weight, water-soluble epoxy resins. These single-step schemes are computationally straightforward and enable rapid network generation. However, they typically result in networks with low degrees of crosslinking. This limitation stems from the fact that real curing is a slow, diffusion-limited process involving local relaxation and molecular rearrangement—effects that are not captured in a single-step simulation. Consequently, it is difficult to achieve experimentally relevant crosslinking degrees (typically $>$80\%) with such methods~\cite{girard1995epoxy,wu2006atomistic}.

 To overcome these limitations, \citet{varshney2008molecular} introduced a multi-step crosslinking protocol incorporating intermediate relaxation steps. This approach enables crosslinking degrees often exceeding 90\% and was successfully used to study the thermal properties of crosslinked EPON-862/DETDA using the CVFF force field, yielding results in good agreement with experiments. \citet{komarov2007highly} proposed a method where monomers are first mapped onto a coarse-grained construct, crosslinked using a Monte Carlo approach, and then back-mapped to an all-atom topology using the OPLS force field. Reactive MD methods have also been explored: \citet{vashisth2018accelerated} used accelerated ReaxFF, while \citet{patil2021reactive} used the Reactive Interface Force Field~\cite{winetrout2024implementing} to simulate the crosslinking process.

The current study employs a multi-step iterative crosslinking scheme, similar to the one proposed by \citet{varshney2008molecular} and implemented in the MAPS\textsuperscript{\texttrademark} software~(\cite{scienomics2025maps}). Variants of this approach have been successfully applied to simulate a wide range of epoxy–crosslinker systems with high degrees of crosslinking~\cite{yang2010study,sahraei2019insights,li2010molecular,li2011molecular,shokuhfar2013effect,liu2023effects,kallivokas2019molecular}. We further propose a modification to the algorithm to restore unreacted crosslinking sites to their original chemical states, avoiding partially modified intermediates—an artifact commonly observed in existing implementations. Details of the algorithm used in this study are provided in Section-\ref{sec:cl}.

The main objective of this paper is to investigate the effect of resin composition on the thermal, structural, and topological properties of the resulting thermosets. To this end, we prepared multiple samples with varying mass fractions of DGEBA and BER while maintaining stoichiometric balance by adjusting the number of monomer molecules accordingly. Details regarding the preparation of the cured thermosets are presented in Section~\ref{sec:SP}. The samples were subjected to equilibration and cooling from \SI{600}{\kelvin} to \SI{100}{\kelvin}, as described in Section~\ref{sec:EC}.

The results and discussion section begin with an analysis of density, which was found to decrease almost linearly with increasing BER content. Subsequently, the coefficients of thermal expansion (CTE) and glass-transition temperatures ($T_g$) were obtained from the cooling simulations. The CTE was observed to increase, while $T_g$ decreased with higher BER content. We subsequently analyzed the free volume and inter-nitrogen path lengths, which indicated that the addition of BER leads to a looser network structure, providing greater free volume. Finally, a discussion is presented that holistically connects the results, highlighting how the addition of BER to DGEBA influences the properties of the crosslinked system.

\section{Preparation of Samples}\label{sec:SP}
\subsection{Simulated Samples}

A total of six crosslinked epoxy samples have been simulated, each comprised of 128 activated Bisphenol A diglycidyl ether~(DGEBA) and/or Butylated epoxy resin~(BER) units(\cref{fig:DGEBA,fig:BER}) along with 64 monomers of activated DETDA~(\cref{fig:DETDA}) as the crosslinker. Here, activation refers to the modification of the epoxide group~(one of which is highlighted in  cyan in \cref{fig:DGEBA}) in the case of DGEBA/BER molecules and the dehydrogenation of the nitrogen atoms in the crosslinker molecules. This activation is done for subsequent crosslinking. For each simulated system, the mass fraction and number of DGEBA and BER monomers is shown in~\cref{table:systems}. The DREIDING~\cite{mayo1990dreiding} force field has been used for the crosslinking and the PCFF~\cite{sun1994ab} force field has been used for production runs. Different force fields have been used for crosslinking and for simulations because the simple nature of the Dreiding force field allows for easy crosslinking, but it has been shown to yield lower densities when compared to experimental data~\cite{sun2018molecular} and, hence, had to be swapped for PCFF during production runs.

\begin{figure}
     \centering
     \setchemfig{atom sep=2em}
     \begin{subfigure}[b]{0.6\textwidth}
         \centering
         \chemfig{(-[1])(-[-3]*6(-=-(-[-3]O-[3]-[-3](-[3]H_2\textcolor{red}{\dot{C}})-HO)=-=))(-[3])(-[-1]*6(-=-(-[-1]O-[1]-[-1]@{e2}C(-[1]\textcolor{red}{\dot{C}}@{H2}H_2)-OH)=-=))}
         \chemmove{
                      \draw[-,
                        fill=cyan,
                        draw=cyan,
                        fill opacity=.1,
                        rounded corners=2pt
                      ]
                        (e2.west) -- ++(0,.8) -| (H2.east)
                        (e2.west) -- ++(0,-.8) -| (H2.east) ;
                    }
         \caption{DGEBA}
         \label{fig:DGEBA}
     \end{subfigure}
     
     \begin{subfigure}[b]{0.6\textwidth}
     \setchemfig{atom sep=1.5em}
         \centering
         \chemfig{(-[1])(-[-3]*6(-=-(-[-3]O-[3]-[-3](-[-3]-[-1]O-[1]-[-1]-[1]-[-1])-[3]O-[-3]-[3](-[3]H_2\textcolor{red}{\dot{C}})-HO)=-=))(-[3])(-[-1]*6(-=-(-[-1]O-[1]-[-1](-[-1]-[-3]O-[3]-[-3]-[3]-[-3])-[1]O-[-1]-[1](-[1]\textcolor{red}{\dot{C}}H_2)-OH)=-=))}
         \caption{BER}
         \label{fig:BER}
     \end{subfigure}

     \begin{subfigure}[b]{0.6\textwidth}
         \centering
         \chemfig{*6(-(-[-2])=(-[-1]\textcolor{blue}{\ddot{N}})-(-[1]-[-1])=-(-[3]-[-3])=)(-[-3]\textcolor{blue}{\ddot{N}})}
         \caption{DETDA}
         \label{fig:DETDA}
     \end{subfigure}
        \caption{Epoxy and crosslinker monomers used to create crosslinked epoxy samples. Carbon atoms that have a free valance electron are colored red and nitrogen atom of the crosslinker with two valance electrons have been colored blue. One of the modified epoxide group of the DGEBA monomer has been highlighted in cyan.}
        \label{fig:monomers}
\end{figure}

\begin{table}
\caption{Various compositions used for the current simulations.}
\begin{ruledtabular}
\begin{tabular}{ccccc}
Sample no.&
\% DGEBA&
no. molecules DGEBA&
\% BER&
no. molecules BER\\
\hline
1&100&128&0&0\\
2&80&113&20&15\\
3&60&93&40&35\\
4&40&70&60&53\\
5&20&40&80&88\\
6&0&0&100&128\\
\end{tabular}
\end{ruledtabular}
\label{table:systems}
\end{table}

The amorphous builder module in the MAPS\textsuperscript{\texttrademark} program was utilized to populate the simulation domain with the monomer molecules. The amorphous builder uses a recoil growth based Monte-Carlo method~\cite{consta1999recoil} in order to optimally populate the domain while minimizing overlaps. The systems were created with an initial density of \SI{1}{\gram\per\centi\meter^3} in a cubic domain with periodic boundary conditions. 

\subsection{Crosslinking}\label{sec:cl}
\begin{figure}
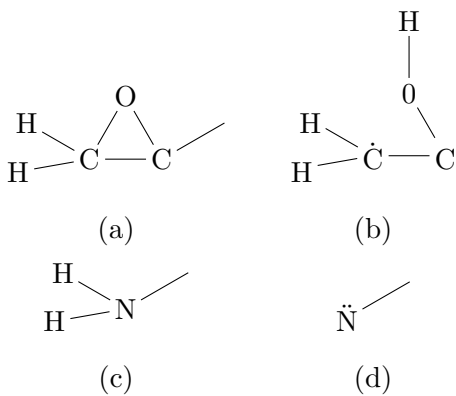

\setchemfig{atom sep=2.5em}
     \centering
     \begin{subfigure}[b]{0.2\textwidth}
         \centering
         \chemfig{C*3([:-30](-[:150]H)(-[:190]H)-C(-[:30])-O-)}
         \caption{}
         \label{fig:epoxide}
     \end{subfigure}    
     \begin{subfigure}[b]{0.2\textwidth}
         \centering
         \chemfig{\dot{C}(-[:150]H)(-[:190]H)(-[:0]C-[:120]0-[:90]H)}
         \caption{}
         \label{fig:m_epoxide}
     \end{subfigure}
     
     \begin{subfigure}[b]{0.2\textwidth}
         \centering
         \chemfig{N(-[:150]H)(-[:190]H)(-[:30])}
         \caption{}
         \label{fig:amin}
     \end{subfigure}
     \begin{subfigure}[b]{0.2\textwidth}
         \centering
         \chemfig{\ddot{N}(-[:30])}
         \caption{}
         \label{fig:m_amine}
     \end{subfigure}
     \caption{Modification of epoxide and amine group for crosslinking. (a)Original epoxide group (b) Modified epoxide group with one free valency (c) Original primary amine group (d) Modified amine group with two free valencies.}
     \label{Fig:activation}
\end{figure}

Once the simulation domain is filled with the monomers, the next step is to crosslink the monomers to form a network. For this purpose the crosslinker module of MAPS\textsuperscript{\texttrademark} program has been utilized. The crosslinker module is based on the algorithm developed by~\citet{varshney2008molecular} and~\citet{li2010molecular}.
Before the crosslinking is started, it is precursory to identify the `reaction centers' i.e., the atoms which can bond with each other to form the crosslinks. It is important that these atoms have the necessary valencies available to form the crosslinks. In our case, the reaction centers are the terminal carbons colored red in the epoxy molecules and the nitrogen atoms colored blue in the crosslinker molecule as seen in~\cref{fig:monomers}. The epoxide group is a three atom ring with an oxygen as seen in~\cref{fig:epoxide}, this has been modified into an open structure(~\cref{fig:m_epoxide}), leaving the terminal carbon with a free valency. Similarly, the nitrogens in the amine group of the DETDA molecule have been dehydrogenated to create two empty valencies~(\cref{fig:m_amine}). Both these modifications are done to create the required free valencies and mimic what happens during a crosslinking reaction. During actual crosslinking, a hydrogen is transferred from the nitrogen in the amine group of DETDA onto the oxygen in the epoxide group, leaving the nitrogen with a free valency. This oxygen breaks the bond with the terminal carbon~(breaking the epoxide ring) to take up this hydrogen from the nitrogen. This leaves the terminal carbon in the epoxide group with a free valency too. This carbon uses the valency to form the crosslinking bond with the nitrogen which also has a free valency. Essentially, we are preempting all the steps in the real crosslinking reaction leaving the terminal carbon atoms in the epoxide group and the nitrogen atoms in the crosslinker with empty valencies. The only remaining step is to form the crosslinking bonds which is taken care by the crosslinking algorithm of the MAPS\textsuperscript{\texttrademark} program.

Once the reaction centers and their free valencies have been specified~(1 for the terminal carbons, 2 for the nitrogens), the next step is to choose the cutoff distance for crosslinking. The cutoff distance determines if a carbon and nitrogen will eventually form a crosslinking bond. There is a certain trial and error involved in choosing the desired cutoff distance. For the samples in this study, the cutoffs are increased from \SI{6}{\angstrom} for the system with pure DGEBA to \SI{8}{\angstrom} for the system with pure BER. Broadly, the cutoff distance required to achieve similar \%crosslinking in the same `time' for different samples is proportional to the average size of the monomers in the sample. Here we define the \%crosslinking as the percentage of carbon atoms that have formed the crosslinking bonds among those that have free valencies. It is to be noted that one nitrogen can potentially bond with two carbon atoms. We have not artificially altered the probability of a primary amine~($NH_2$) and a secondary amine~($NH$) forming a crosslink, and kept them both the same. That being said, given that a secondary amine may not have the same mobility as a primary amine, the dynamics of the system may result in alteration of said probabilities.

\begin{figure}
  \includegraphics[scale = 0.7]{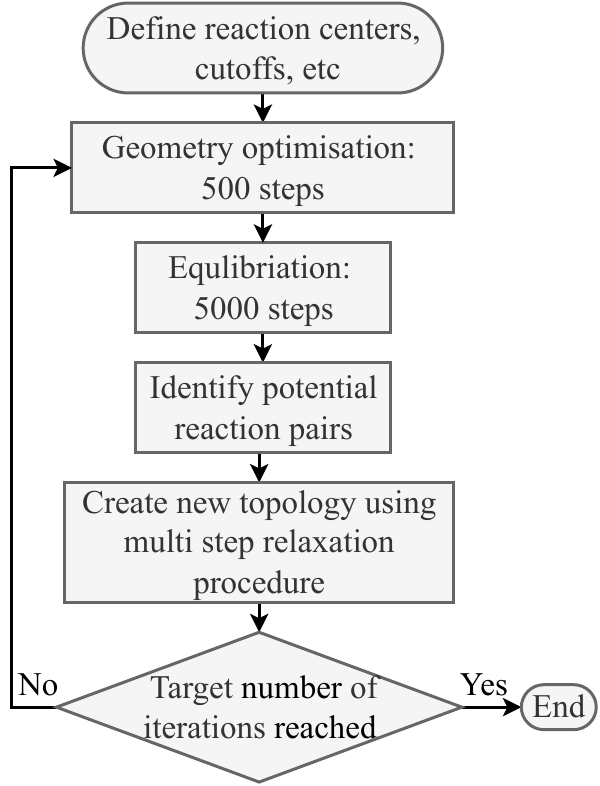}
  \caption{Flowchart representation of the crosslinking algorithm.}
  \label{fig:cl_algo}
\end{figure}

The crosslinker uses an iterative multi-step relaxation scheme as seen in~\cref{fig:cl_algo}. We have previously defined the reaction centers and cutoffs, the next step is to initiate the iterative crosslinker algorithm. Every iteration starts with a 500 steps geometry optimization followed by a 5000 time steps equilibration. This is followed by identification of reaction pairs, which are essentially carbons and nitrogens having free valencies within the cutoff distance. If reaction pairs are found, new bonds are added between them in a multi-step relaxation process. Multi-step relaxation process means that the bond force constants of the newly formed crosslink bonds are gradually increased in 20\% increments with a 1000 timestep equilibration at each relaxation step to gradually relax the topology and prevent instabilities. All equilibration steps in this crosslinking process are MD simulation runs in an NVT ensemble at \SI{500}{\kelvin} with a timestep of \SI{1}{\femto\second}.

\begin{figure}
  \includegraphics[scale = 0.45]{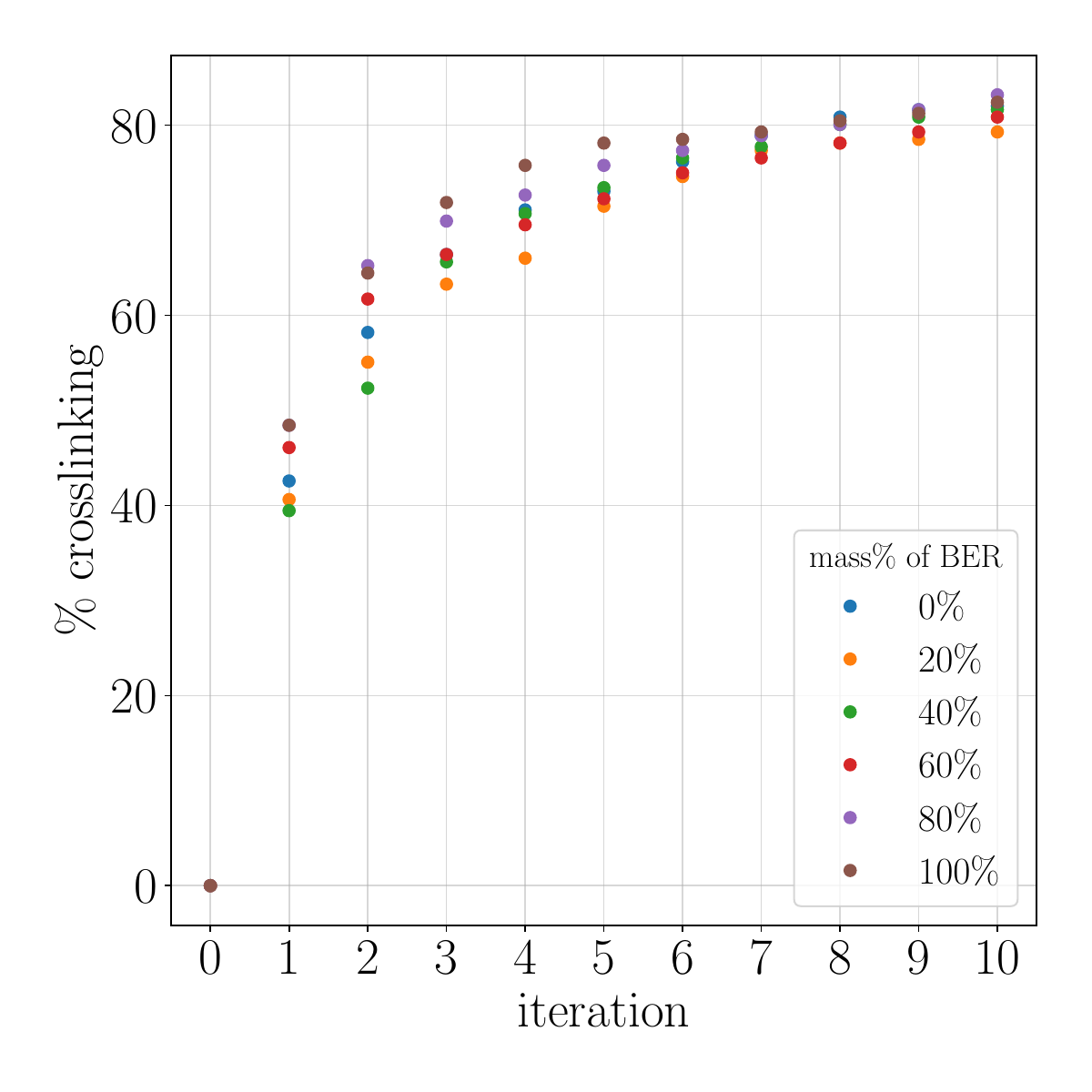}
  \caption{Evolution of the percentage of crosslinking as the iterations of the crosslinking algorithm progress.}
  \label{fig:cl}
\end{figure}

The cutoffs have been chosen to achieve around 80\% crosslinking upon completion of 10 iterations of the algorithm.~\Cref{fig:cl} shows the progression of the \%crosslinking as the iterations progress. We see that each simulated system, indicated by the mass fraction of BER, follows a similar path and reaches around 80\% crosslinking by the end. As the \%crosslinking increased, the mobility of the molecules reduces and, as a consequence, is harder to rearrange the topology sufficiently to form new crosslinks, hence the \%crosslinking begins to level off at 80\%. We have chosen this particular value as our target crosslinking, as the computational cost of achieving higher crosslinking percentage increases steeply beyond this value. While 80\% is a little lower than experimentally realized \%crosslinking, which is closer to 90\%(\cite{varshney2008molecular}), most new bonds that form beyond 80\% crosslinking are intramolecular~(\cite{varshney2008molecular}) and have a limited effect on the topology or dynamics of the system. 

\subsection{Uncrosslinked Sites}

While crosslinking ensures that most of the valencies in the system are filled, there are still few epoxide groups and/or amine groups that have not formed/fully formed crosslinks, and, hence, are still having some empty valencies. In case of nitrogens, this is simple enough to deal with, as all we need to do is to saturate the nitrogens with hydrogens. But as for the carbons which have not participated in crosslinking, it is necessary to convert them back into the epoxide group~(\cref{fig:epoxide}). To achieve this, we have reused the crosslinker module and the process is as follows;
\begin{enumerate}
    \item All the oxygen atoms in the system are dehydrogenated and labeled as reaction centers.
    \item The terminal carbons which have not yet formed crosslink sites are also labeled as reaction centers.
    \item The crosslinker is asked to form bonds between said carbon and oxygen atoms with a cutoff of \SI{2.6}{\angstrom} which was found to be the appropriate cutoff to form bonds between carbons and oxygens to complete the epoxide rings, but not so high as to form unwanted bonds.
    \item Since every labeled carbon has at least one oxygen atom in a \SI{2.6}{\angstrom} radius, the crosslinker completes the job in a single iteration.
    \item All the remaining oxygens are saturated with hydrogens.
\end{enumerate}

\begin{figure}
    \centering
    \includegraphics[width=0.75\linewidth]{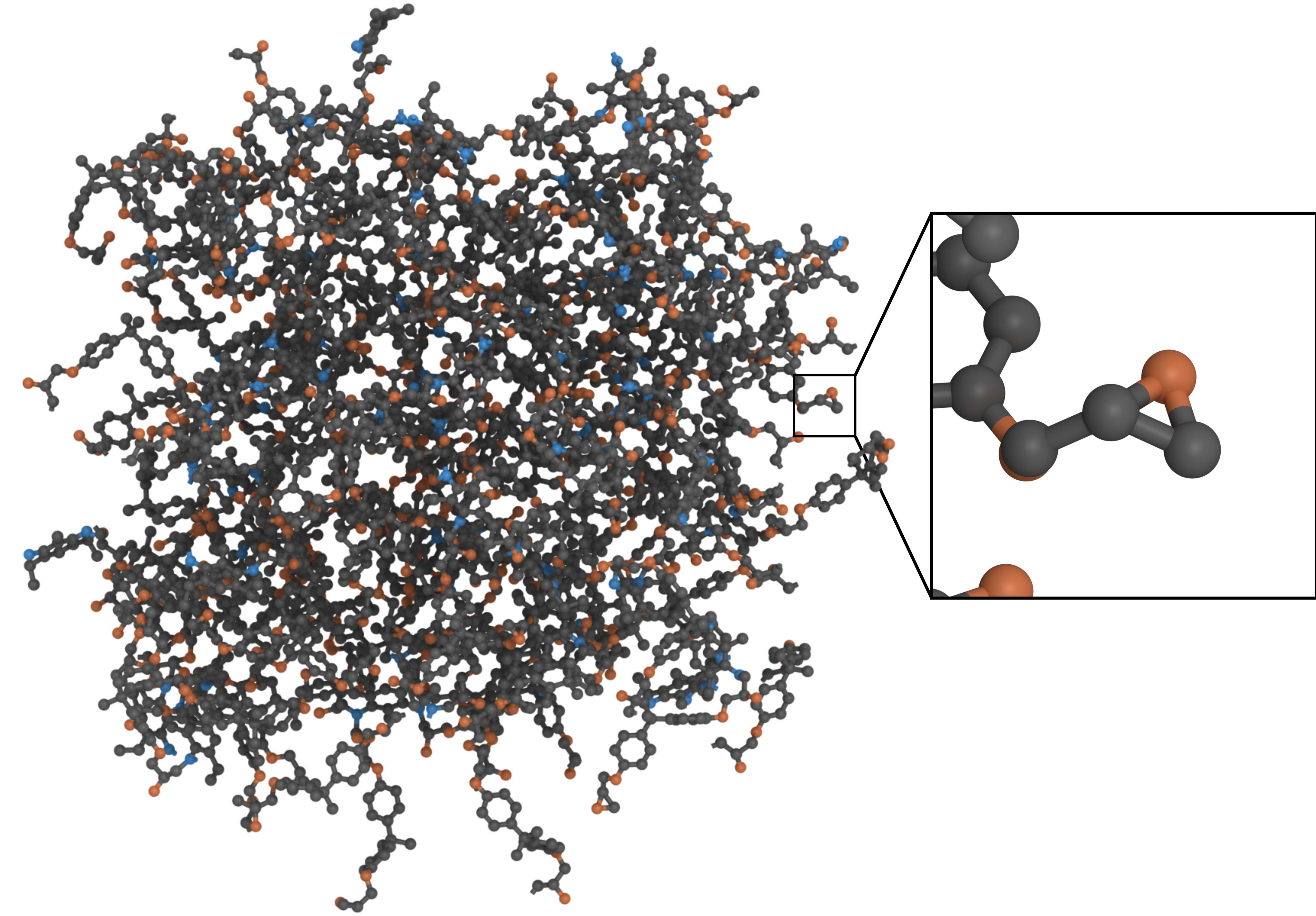}
    \caption{Snapshot of the crosslinked system~(system no. 1, see~\cref{table:systems}) with a zoomed in view of a refurbished epoxide group. The hydrogen atoms have not been displayed for the the sake of visual simplicity. Carbon atoms have been colored dark grey, oxygen atoms have been colored orange and nitrogen atoms have been colored blue.}
    \label{fig:sys1_rend}
\end{figure}

This process returns the correct topology for the uncrosslinked sites.~\cref{fig:sys1_rend} shows the sample with pure DGEBA~(system-1 in \cref{table:systems}), with a zoomed in view of one of the 'reconstructed' epoxide group.

\section{Equilibration and Cooling}\label{sec:EC}

The crosslinked samples have been subjected to a \SI{1000}{\pico\second} NPT equilibration simulation with a time step of \SI{1}{\femto\second} at \SI{600}{\kelvin} and 1 atmosphere pressure in order to relax the system. The thermostat and the barostat have a damping coefficient of \SI{10}{\femto\second} and \SI{350}{\femto\second} respectively. The LAMMPS~(\cite{LAMMPS}) program was utilized to run these simulations. As mentioned previously, the PCFF force field was implemented to simulate the system post crosslinking. Upon completion of the equilibration, the systems were cooled from \SI{600}{\kelvin} to \SI{100}{\kelvin} at a rate of \SI{0.01}{\kelvin\per\pico\second} in order to extract the glass-transition temperature. 

\begin{figure}
     \centering
     \begin{subfigure}[b]{0.45\textwidth}
         \centering
         \includegraphics[width=\textwidth]{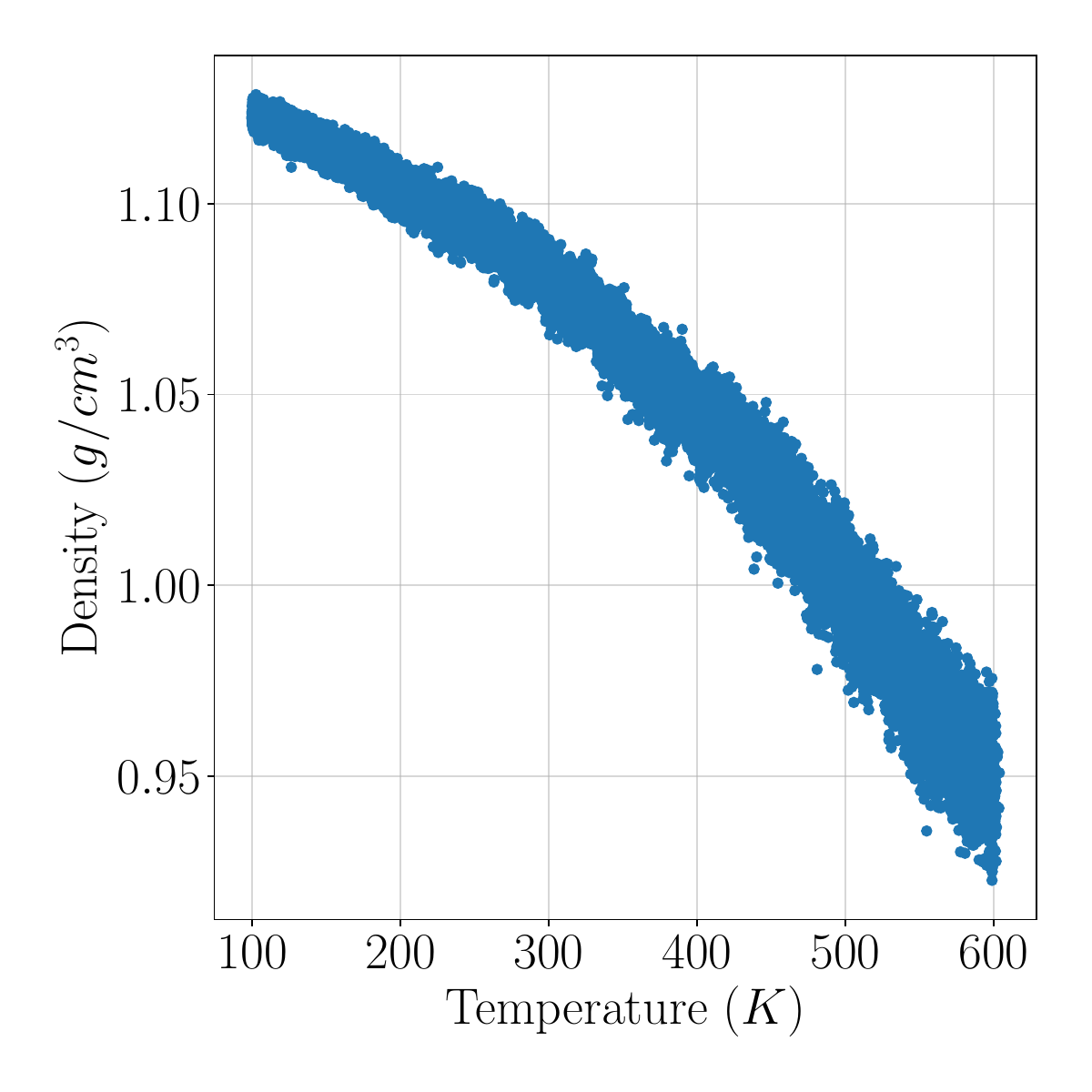}
         \caption{Raw data}
         \label{fig:raw data}
     \end{subfigure}
     \begin{subfigure}[b]{0.45\textwidth}
         \centering
         \includegraphics[width=\textwidth]{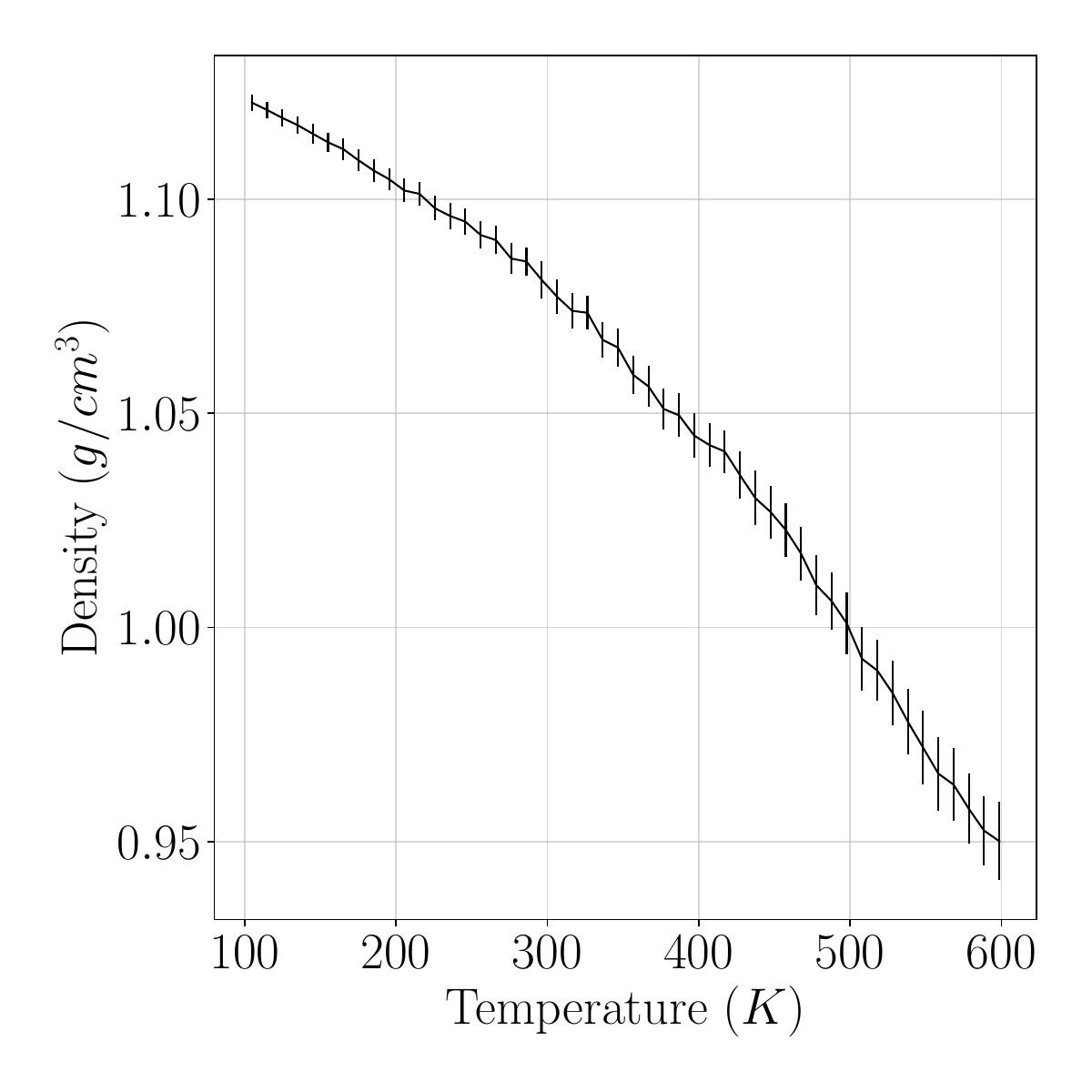}
         \caption{Binned data}
         \label{fig:bin_data}
     \end{subfigure}
        \caption{Temperature-density data obtained from cooling simulation of the samples containing pure DGEBA. In the plot showing the binned data, the error bars showcase the standard deviation of the data within the bin.}
        \label{fig:data}
\end{figure}

We have utilized the shift in the slope of the density-temperature plot to identify glass-transition temperature. ~\cref{fig:raw data} shows the raw density-temperature data from the cooling simulation of one of the system~(system no. 1, \cref{table:systems}). While we can see that there is a point somewhere in the middle where the slope of the plot changes, it is hard to pinpoint a specific temperature.

\begin{figure}
  \includegraphics[scale = 0.45]{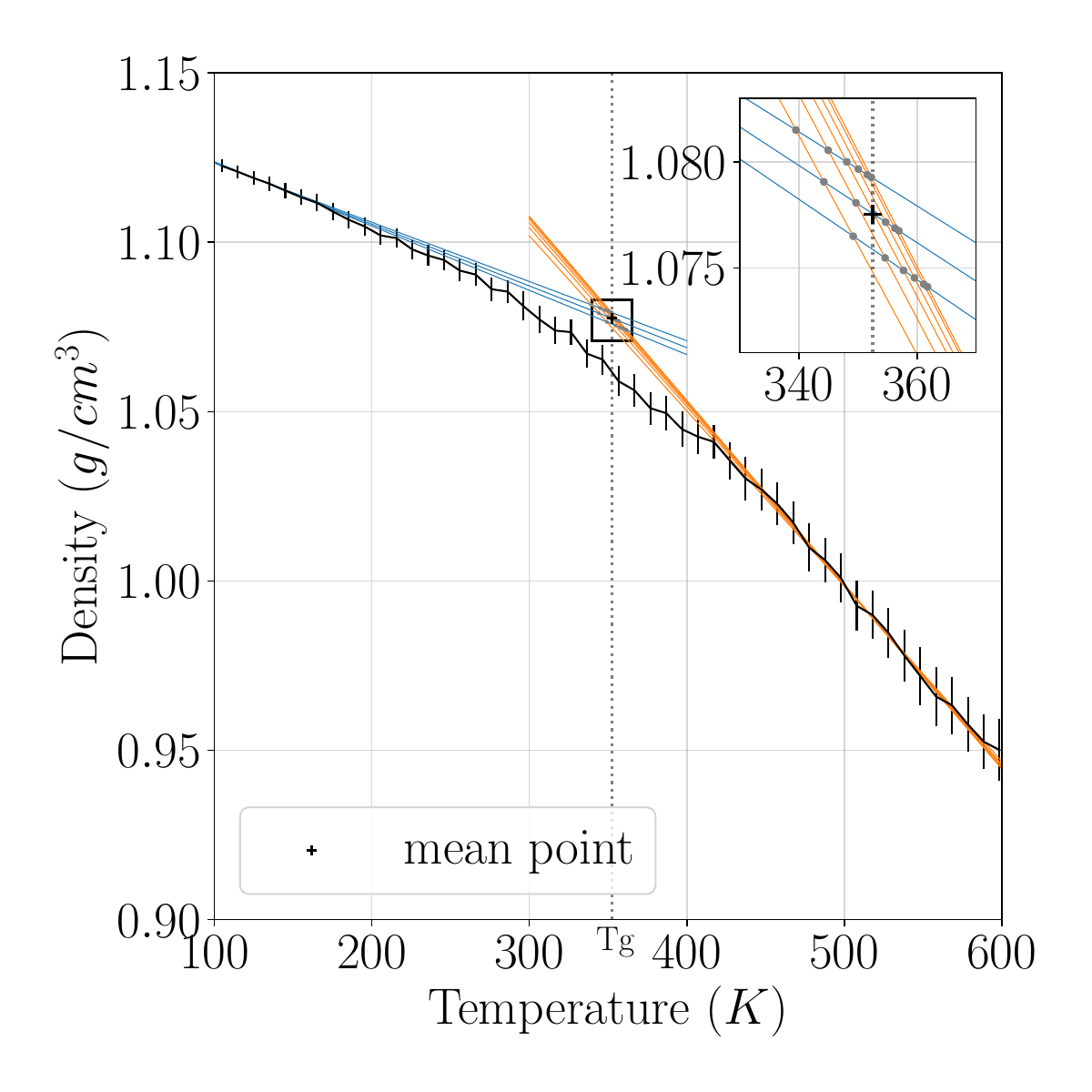}
  \caption{Binned temperature-density data along with the lines projecting from high temperature extreme and low temperature extreme. The region near the point of intersection of the lines has been zoomed in at the top right corner of the plot. The black '+' represents the average value of the points of intersection.}
  \label{fig:bdt}
\end{figure}

In order to compute a good estimate of the glass-transition temperature, we first divide the data into 50 partitions of equally spaced temperature bins and compute the average of the density in every bin, essentially creating a mid-point moving average with a window size of \SI{10}{\kelvin}, see ~\cref{fig:bin_data}. The next step is to find a point of where the slope of this data changes, the most common way to do that is to find the point of intersection of the lines that coincided with the data on the either extremes. In order to do so, we have isolated 6 windows of data on the lower temperature extreme and 18 windows on the higher temperature extreme. Now of 6 windows in the lower temperature extreme, we have taken every 4 successive points, that is, if the bins are $b_1$, $b_2$, $b_3$, $b_4$, $b_5$ and $b_6$, we have taken ($b_1$, $b_2$, $b_3$, $b_4$), ($b_2$, $b_3$, $b_4$, $b_5$) and ($b_3$, $b_4$, $b_5$, $b_6$) and constructed linear fits of these sets, which gives us three lines for the lower temperature range. Similarly for the 18 bins in the higher temperature range, we have chosen every 12 successive bins which gives us 7 lines in the higher temperature range. Now the points of intersection of all the lines in the lower temperature range and the higher temperature range are computed, which results in 21 points of intersection which as seen in~\cref{fig:bdt}. The mean of these 21 points is identified as the glass-transition temperature with the standard deviation of the x-coordinate of these points giving the error in calculating the glass-transition temperature.

\section{Results and Discussion}\label{sec:RD}

Subsections A through E report the results in a factual manner, starting with the system density, followed by thermal properties such as the coefficient of thermal expansion and glass-transition temperature. This is followed by examination of the atomic mobility. Subsequently moving on to volumetric and topological analyses including Voronoi atomic volumes, nitrogen-nitrogen path lengths, and the confinement index—a novel metric developed in this study to quantify local atomic packing. Subsection F then offers a detailed discussion and interpretation of these results, connecting the observed trends to the underlying molecular structure and network topology.

Together, these results provide insight into how the addition of BER influences the structural integrity and thermal behavior of the crosslinked epoxy networks. Special attention is given to understanding the mechanisms driving the reduction in glass-transition temperature and the implications of network loosening on material performance.

\subsection{Density}
\begin{figure}
  \includegraphics[scale = 0.45]{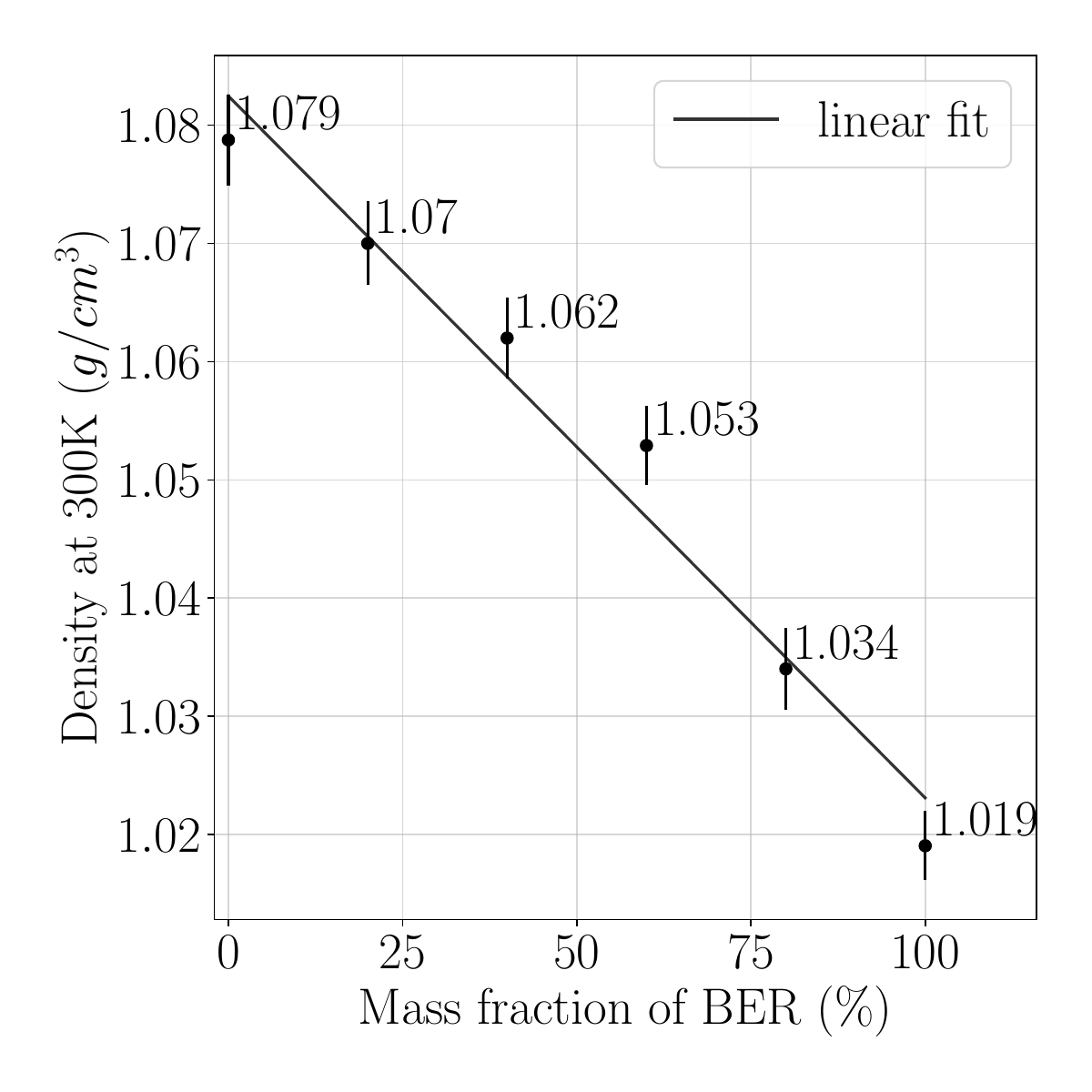}
  \caption{Average density of the crosslinked samples at \SI{300}{\kelvin} as a function of the mass fraction of BER.}
  \label{fig:density}
\end{figure}

We begin analyzing the simulations by looking at the density of the simulated crosslinked systems and comparing them with available literature.~\cref{fig:density} shows the density of the systems sampled at $300\pm2K$ as a function of mass fraction percentage of BER. The density of the system comprising pure DGEBA is approximately $\SI{1.079}{\gram\per\centi\metre^3}$. \citet{varshney2008molecular} simulated DGEBF~(EPON\textsuperscript{TM}-862) crosslinked with DETDA, and reported a density of $\SI{1.12}{\gram\per\centi\metre^3}$ which is very close to the density we obtained. DGEBF is very similar to DGEBA except for the central carbon which is methylated in case of DGEBA. More recently \citet{jeyranpour2015comparative} simulated crosslinked DGEBA/DETDA and reported system density of $\SI{1.123}{\gram\per\centi\metre^3}$ at 75\% crosslinking using the COMPASS force field. This is again very close to the value we obtained with a difference of less than 4\%. We were unable to find any literature on crosslinked systems comprising of BER. There is almost linear trend of reduction in density with the addition of BER. On average, we observe approximately 1\% reduction in density with every 20\% increase in mass fraction of BER in the epoxy. In coming Subsections, the cause behind this reduction in density is explored.

\subsection{Thermal properties}

\subsubsection{Coefficient of Thermal Expansion}

The coefficient of thermal expansion~(CTE) can be easily calculated from the temperature-density data~(\cref{fig:raw data}) by measuring the slope of the temperature-density plot,

\begin{equation}
    \alpha = -\frac{1}{\rho} \cdot \frac{d\rho}{dT}
    \label{eq:cte}
\end{equation}

where $\alpha$ is the CTE, $\rho$ is the density and $T$ is the temperature.

\begin{figure}
     \centering
         \centering
         \includegraphics[scale = 0.45]{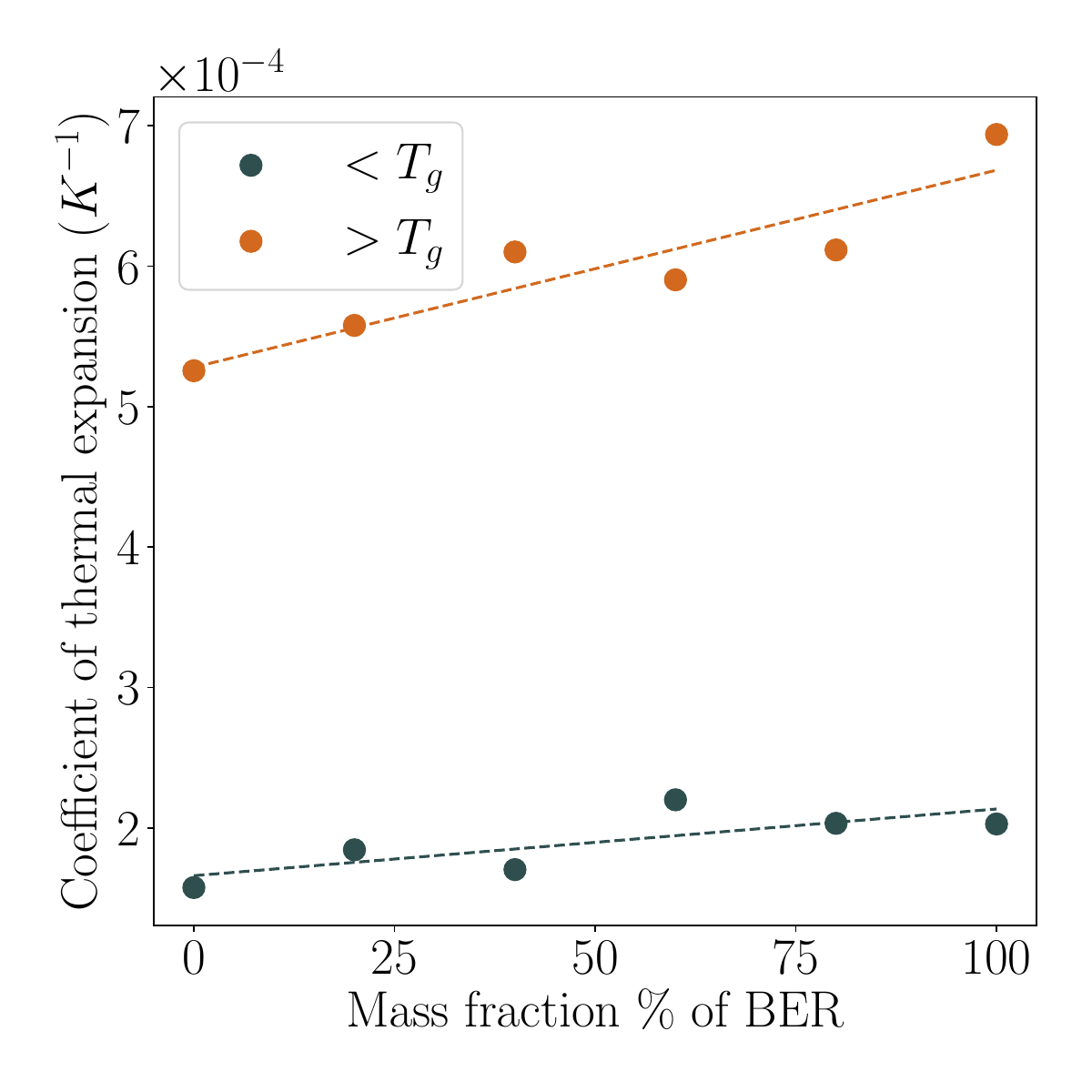}
         \label{fig:cte_brittle}
        \caption{Evolution of the coefficient of thermal expansion~(CTE) as a function of the mass fraction of BER in the samples both above and below $T_g$. The dashed lines represent linear-fitting lines.}
        \label{fig:cte}
\end{figure}

The CTE is computed both in the solid state, i.e., below the glass-transition temperature~(from \SI{100}{\kelvin} to \SI{150}{\kelvin}) and in the melt state above the glass-transition temperature~(from \SI{500}{\kelvin} to \SI{600}{\kelvin}) using a finite difference scheme. \cref{fig:cte} shows the evolution of CTE with the addition of BER to samples both in solid and melt state. 
Both above and below $T_g$, we see a positive correlation between CTE and mass fraction of BER. 

The slope of the linear fit in the melt state is $\mathrm{1.4e}{-6}$ and the slope of the linear fit in the solid state is $\mathrm{5.3e}{-7}$. While there is an increase in the coefficient of thermal expansion of the samples in both the melt and solid states, the rate of increase in the melt state is almost three times the rate of increase in the solid state. 

\subsubsection{Glass-Transition Temperature}
\begin{figure}
  \includegraphics[scale = 0.45]{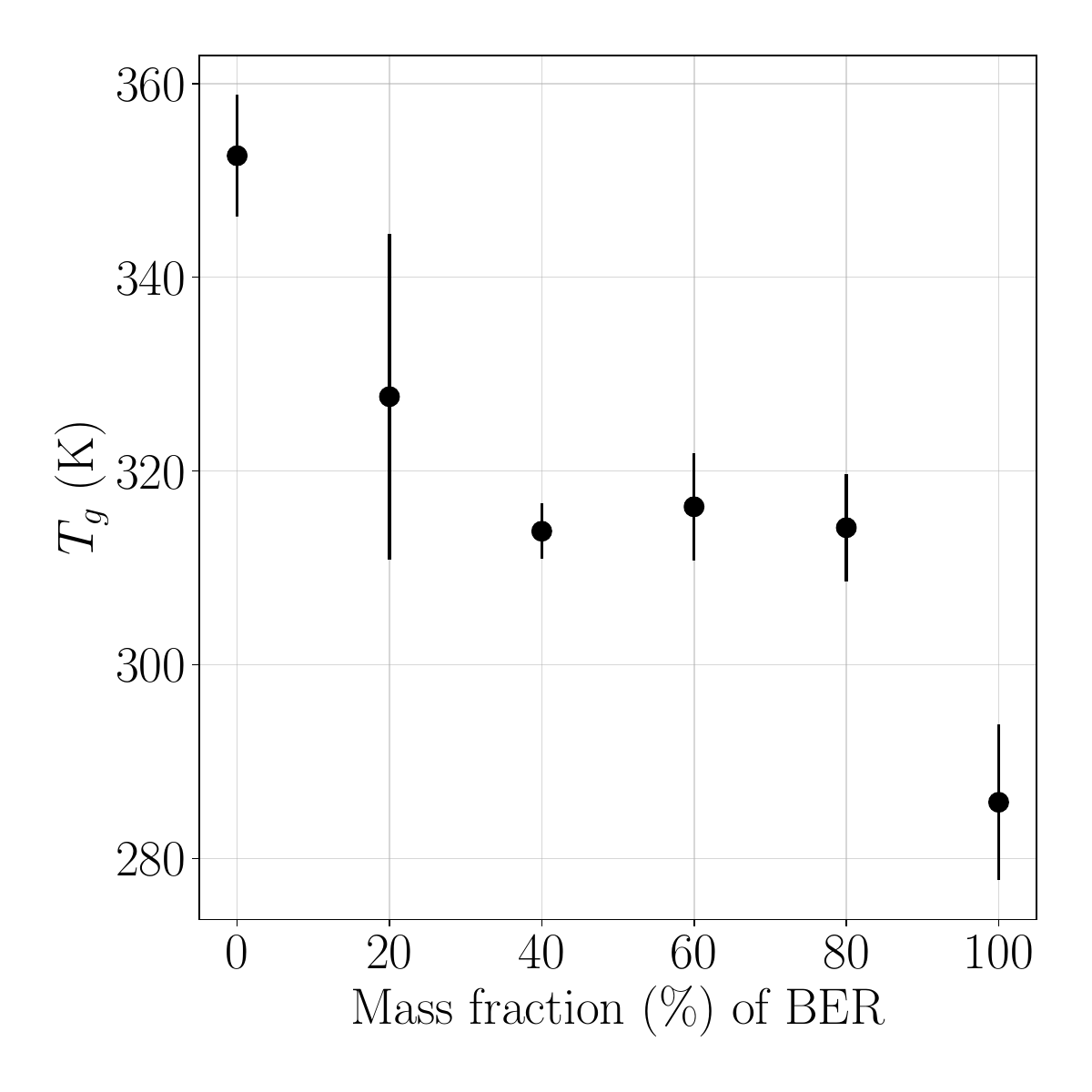}
  \caption{glass-transition temperature of the simulated samples versus the mass fraction of BER}
  \label{fig:tg_comp}
\end{figure}

Next we study the effect of addition of BER on the glass-transition temperature of the crosslinked systems. The method for extracting the glass-transition temperature has been described previously in Section~\ref{sec:cl}. \cref{fig:tg_comp} shows glass-transition temperatures of the samples as a function of the mass fraction of BER. 

While there is an overall reduction in the glass-transition temperature of the samples, the reduction is not linear. A sharp decrease between 0\% and 40\% mass fraction of BER followed by a level-off and the further reduction between 80\% and 100\% is observed. Regarding the range of the glass-transition temperatures, it can be seen that the sample with pure DGEBA has a $T_g$ of \SI{352}{\kelvin} while the sample with pure BER has a $T_g$ of \SI{288}{\kelvin}. This large difference of more than \SI{60}{\kelvin} is addressed in the next section where we study the topological and structural effects of addition of BER to the samples.

\subsection{Atomic Mobility}

The next characteristic we examine is atomic mobility, which is closely linked to the thermal behavior of the system. To quantify this, we computed the root mean square displacement (RMSD) of atoms. RMSD represents the root mean square of the net translational displacement of atoms from a reference configuration and serves as a robust measure of their dynamic freedom within the network.

\begin{figure}
     \centering
     \begin{subfigure}[b]{0.45\textwidth}
         \centering
         \includegraphics[width=\textwidth]{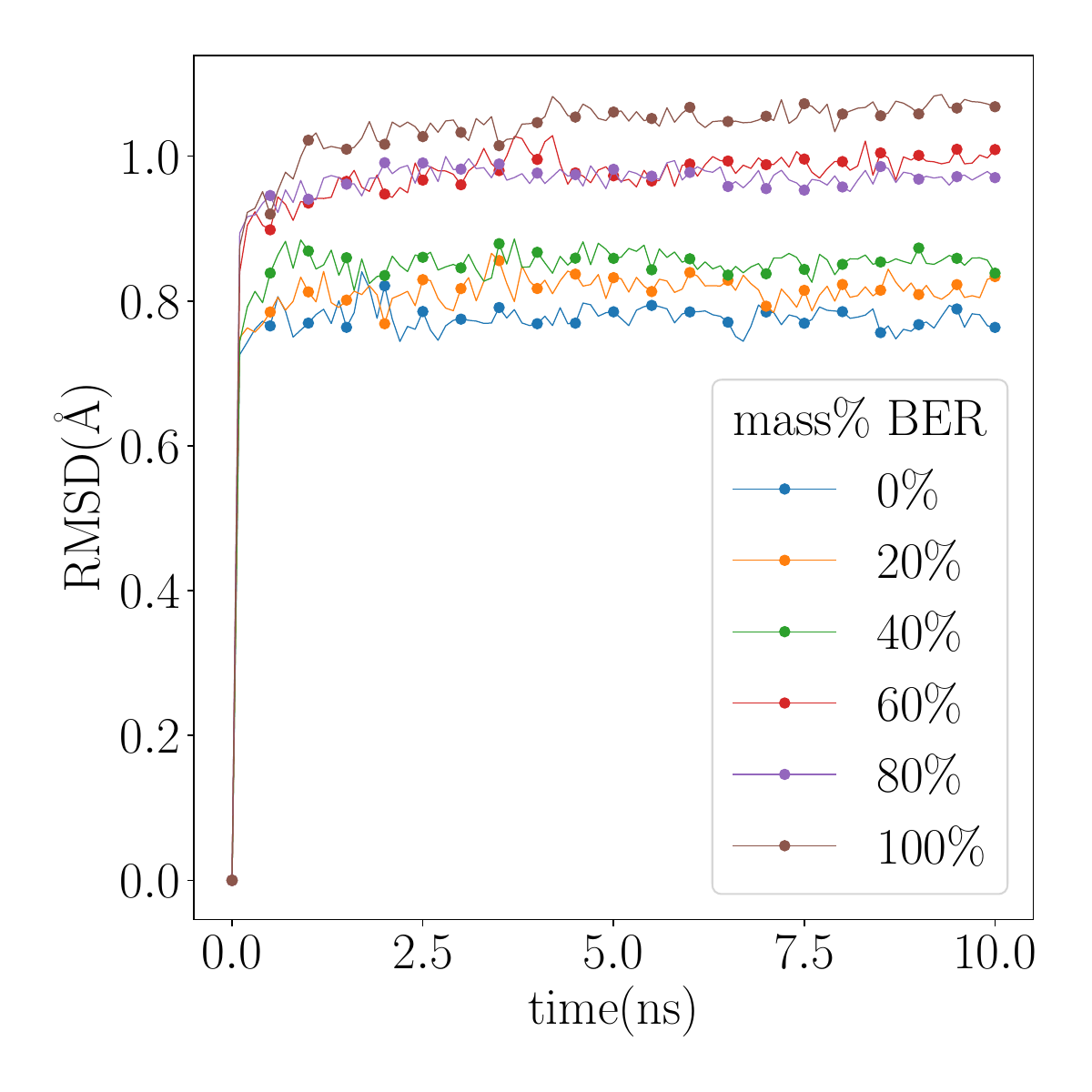}
         \caption{Solid state}
         \label{fig:rmsd_brittle}
     \end{subfigure}
     \begin{subfigure}[b]{0.45\textwidth}
         \centering
         \includegraphics[width=\textwidth]{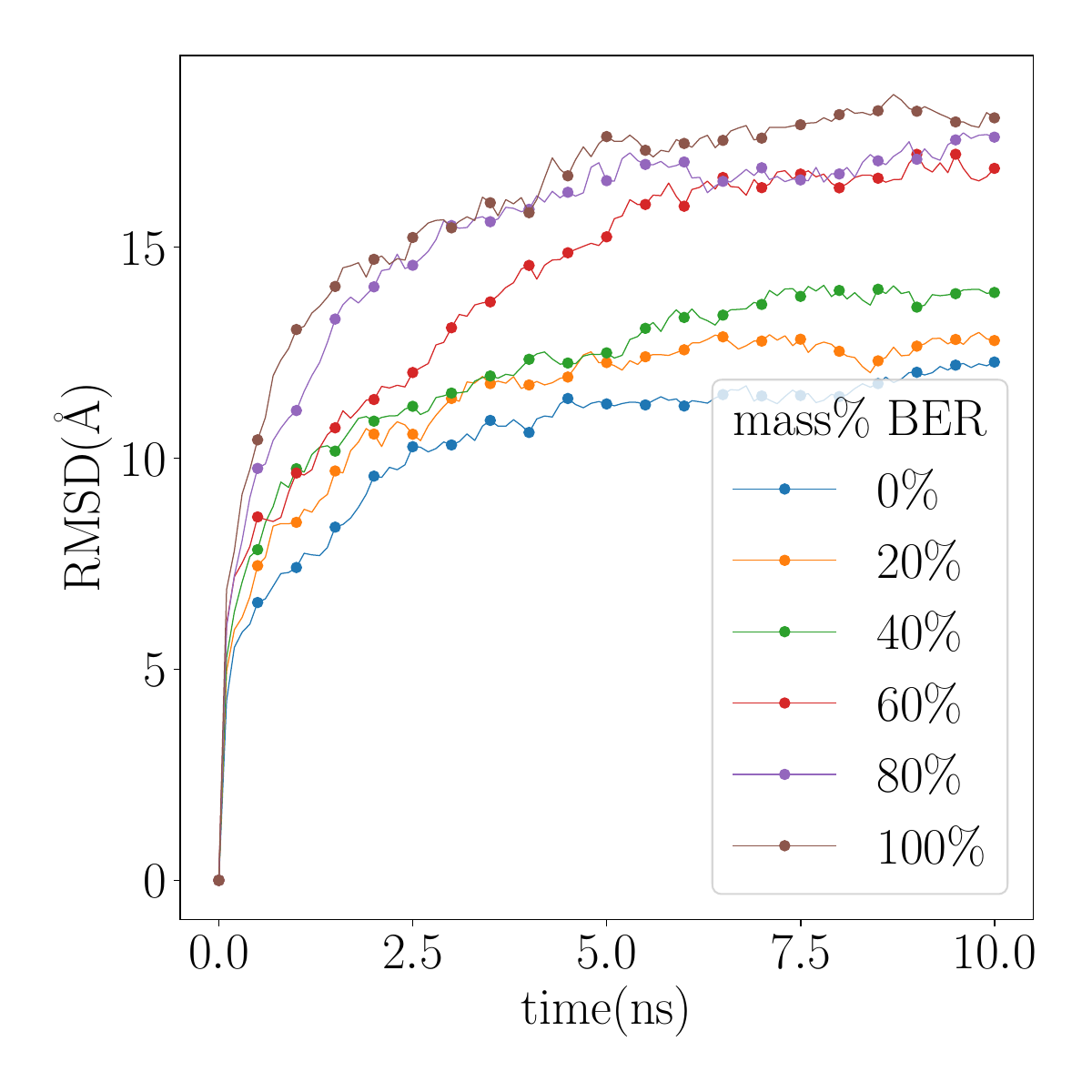}
         \caption{Melt state}
         \label{fig:rmsd_soft}
     \end{subfigure}
        \caption{Root mean square displacement of the backbone atoms~(C, O and N) of the crosslinked system subjected to cooling in~(a) solid state~(from \SI{200}{\kelvin} to \SI{100}{\kelvin}) and in the~(b) melt state~(from \SI{600}{\kelvin} and \SI{500}{\kelvin})}
        \label{fig:rmsd}
\end{figure}

For this analysis, we leveraged the same cooling simulations used in the determination of the glass-transition temperature ($T_g$). The systems were cooled from \SI{600}{\kelvin} to \SI{100}{\kelvin} at a uniform rate of \SI{0.01}{\kelvin\per\pico\second} which results in a total simulation time of \SI{50}{\nano\second}. Of these \SI{50}{\nano\second}, RMSD was evaluated over two distinct \SI{10}{\nano\second} time intervals: one above $T_g$, spanning temperatures from \SI{600}{\kelvin} to \SI{500}{\kelvin}, and the other below $T_g$, from \SI{200}{\kelvin} to \SI{100}{\kelvin}. \cref{fig:rmsd} shows the evolution of the RMSD over the intervals above and below $T_g$ for all the samples. We observe RMSD leveling off in all the samples both above and below $T_g$. The only difference is, in the solid state, the leveling off is instant, and the value at which RMSD levels off is small~($<$ 1.2\AA). This is typically seen when the atomic movement is primarily governed by bond level vibrations. Where as in the melt state, we observe RMSD leveling off more gradually and at much higher values~($\>$ 10\AA). This is a characteristic of dominant molecular level rearrangement which happens much slowly than bond-level vibrations.

\begin{figure}
         \centering
         \includegraphics[scale = 0.45]{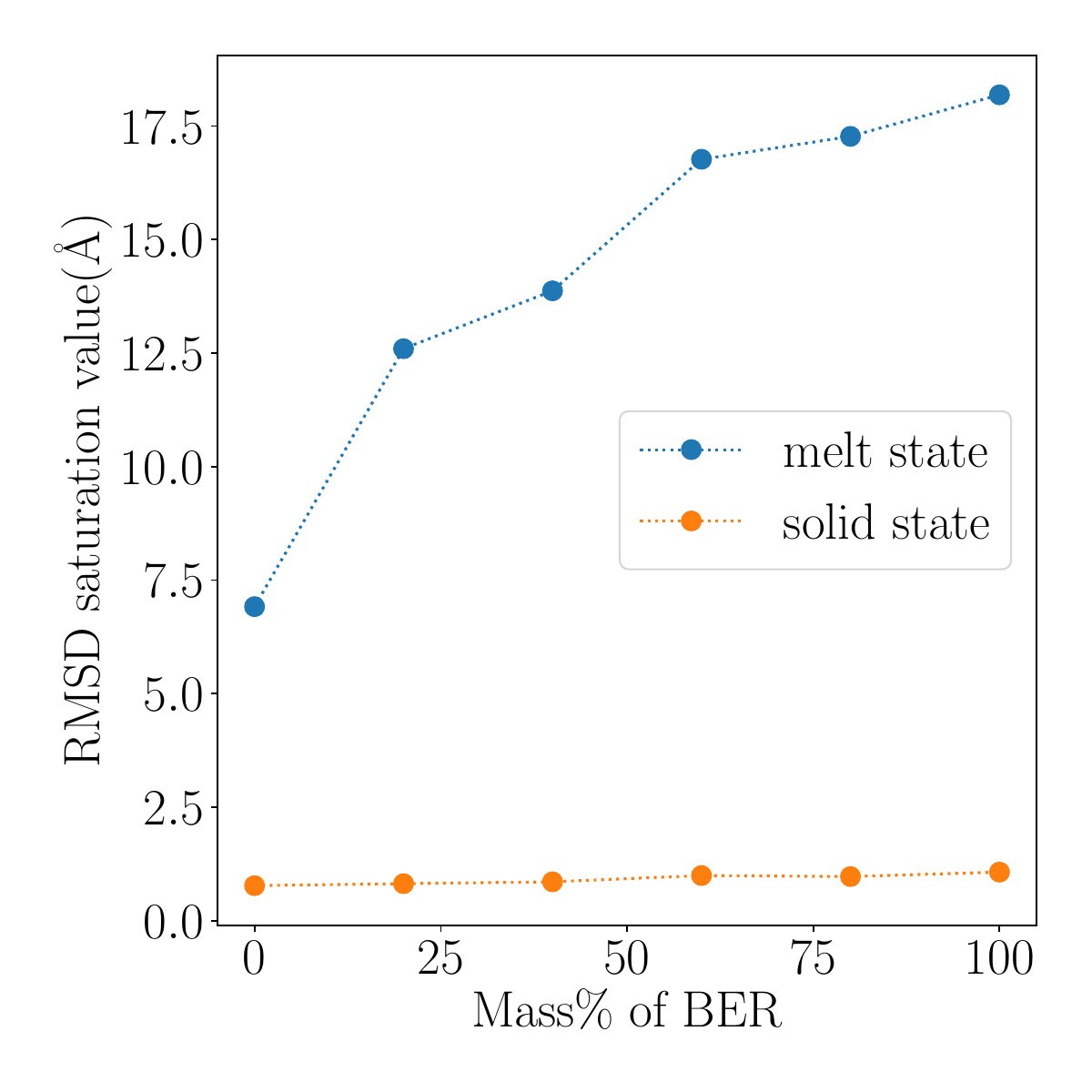}
         \label{fig:rmsd_m_brittle}
        \caption{Steady state RMSD~(average of the RMSD from \SI{8}{\nano\second} \SI{10}{\nano\second} in~\cref{fig:rmsd}) that the backbone atoms reach both in solid state and in melt state.}
        \label{fig:rmsd_mean}
\end{figure}

Notably, the addition of BER leads to an increase in the plateau value of RMSD, particularly in the melt state. To quantify this effect, we extracted the RMSD over the last \SI{2}{\nano\second} of each interval and computed the average. We call this the 'RMSD saturation value'. \cref{fig:rmsd_mean} shows the RMSD saturation value as a function of BER mass fraction. In the melt state, a clear and substantial increase in RMSD saturation value is observed with increasing BER content. While there is also a very small increase in RMSD saturation value in the solid state, it is almost negligent when compared to the melt state. 

\subsection{Atomic volume analysis}

\begin{figure}
     \centering
     \begin{subfigure}[b]{0.45\textwidth}
         \centering         \includegraphics[width=\textwidth]{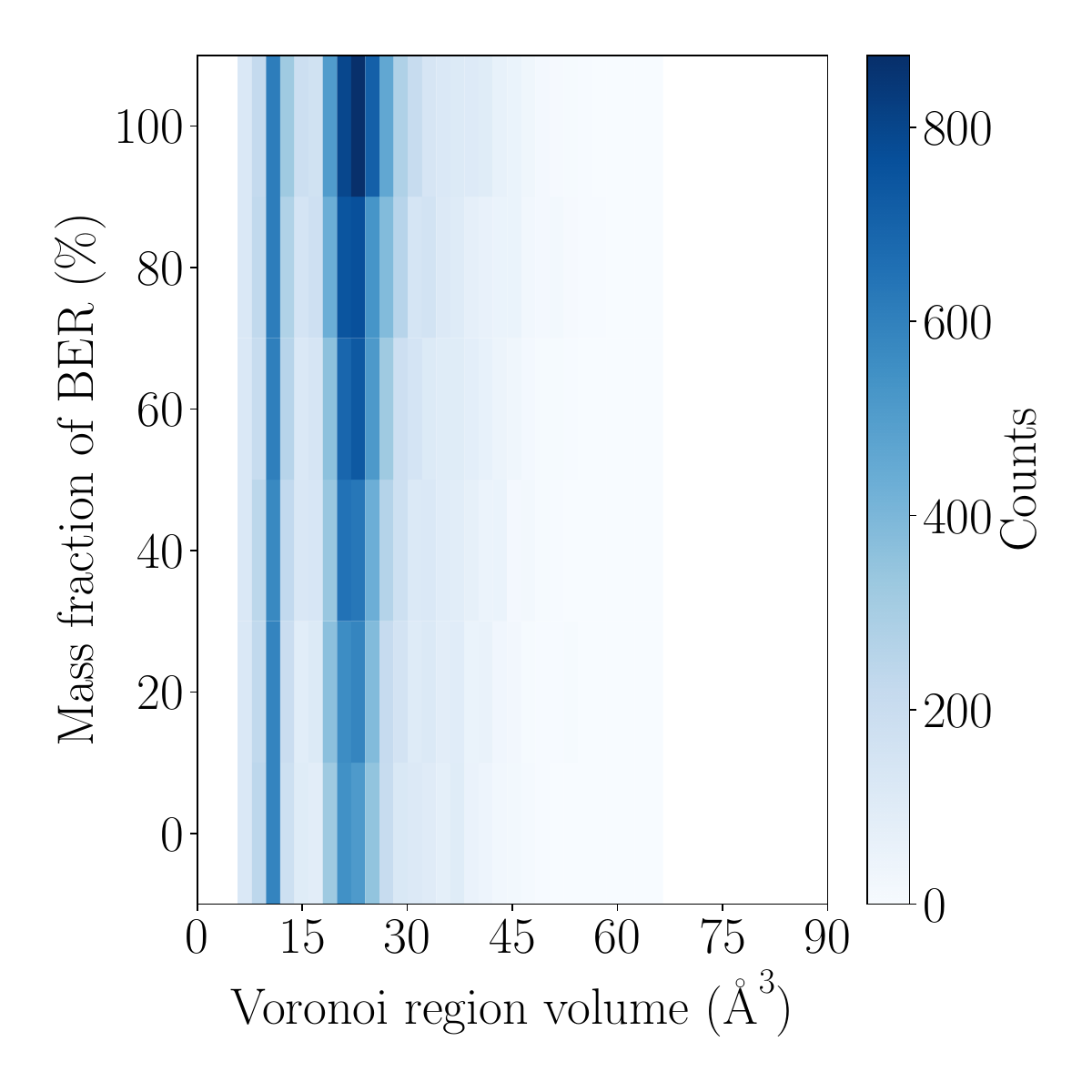}
         \caption{}
         \label{fig:avm_2dhist}
     \end{subfigure}
     \begin{subfigure}[b]{0.45\textwidth}
         \centering
         \includegraphics[width=\textwidth]{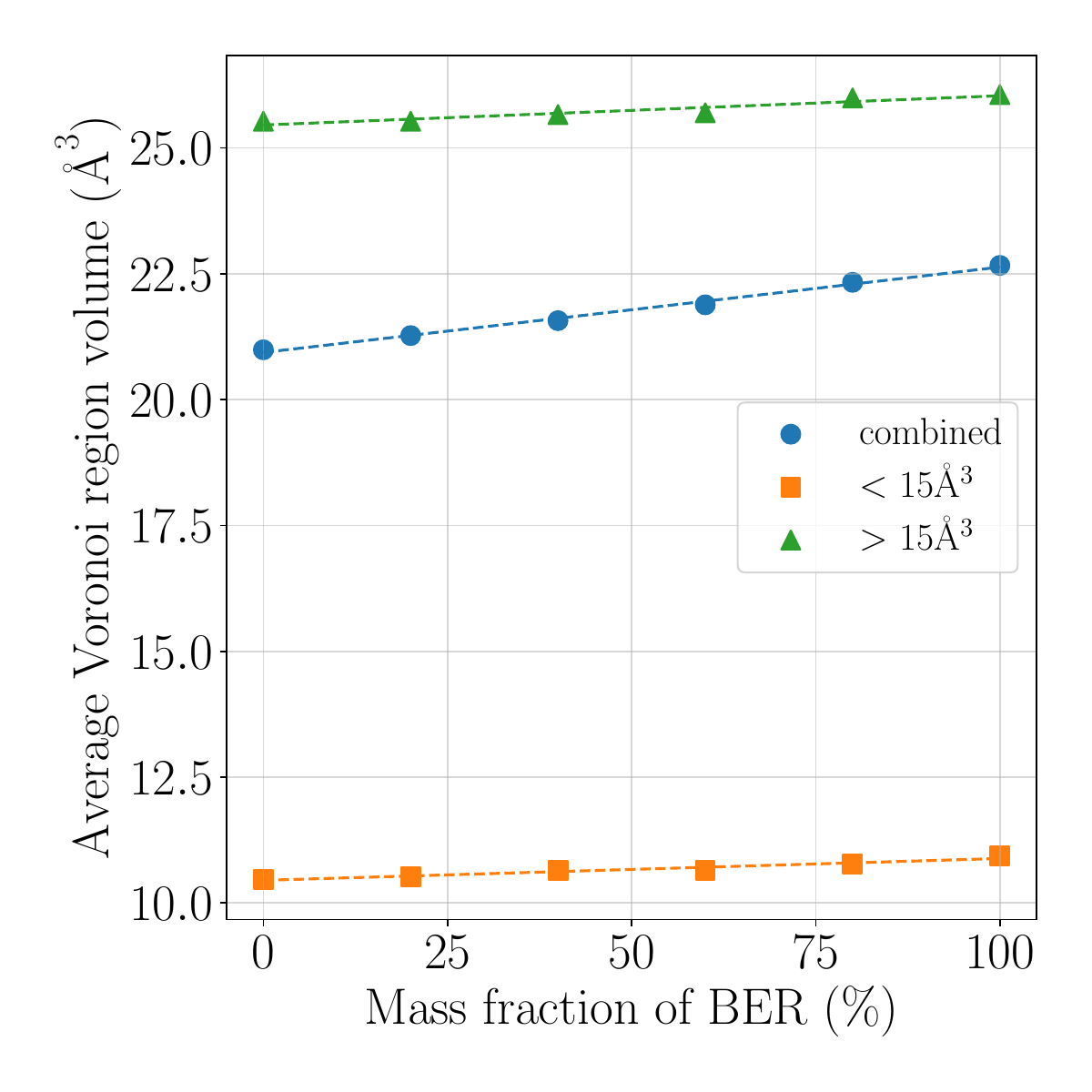}
         \caption{}
         \label{fig:avmmean}
     \end{subfigure}
        \caption{(a) Two dimensional histogram representation of the distribution of atomic volumes across all samples at \SI{300}{\kelvin}. (b) Average atomic volume versus the mass fraction of BER along with average atomic volume of atoms with volume greater than 15 $\textrm{\AA}^3$ and with volume less than 15 $\textrm{\AA}^3$ at \SI{300}{\kelvin}.}
        \label{fig:avm_hm}
\end{figure}

To estimate the local volume available to each atom—referred to here as the atomic volume—we performed a Voronoi tessellation~\cite{Voronoi1908} of the atomic configuration~(excluding hydrogen atoms) and analyzed the resulting Voronoi polyhedra. We used Ovito software~\cite{ovito} where we first removed all the hydrogen atoms from the system. Subsequently, we generated the Voronoi tesselation and extracted the volumes of the Voronoi cells. \cref{fig:avm_2dhist} showcases the histograms of the volumes of the Voronoi cells at 300K. 

\begin{figure}
  \includegraphics[scale = 0.15]{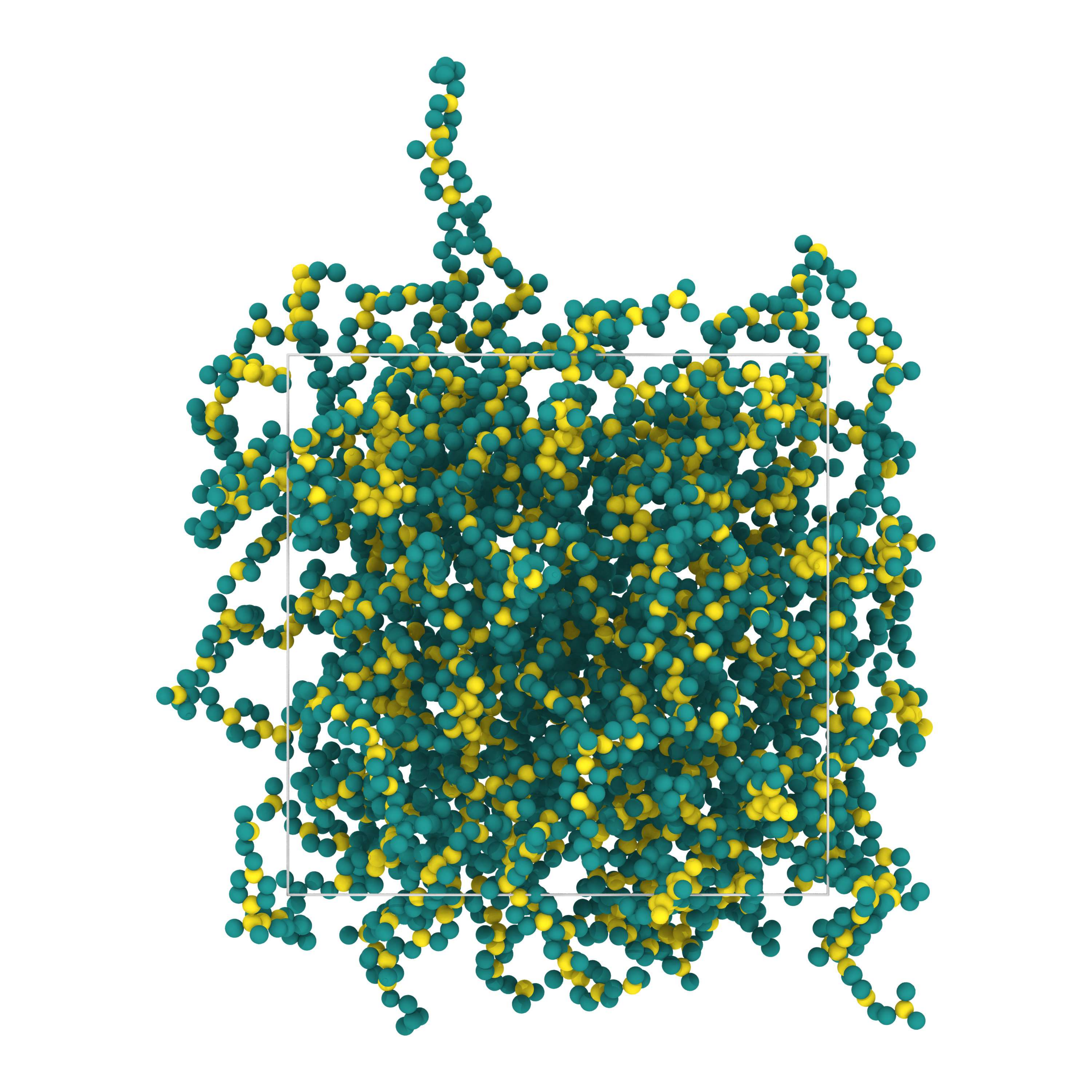}
  \caption{A snapshot of the one of the crosslinked samples containing pure DGEBA with the colors representing the atomic volume, green is above $15\textrm{\AA}^3$ and yellow is below $15\textrm{\AA}^3$.}
  \label{fig:render_lf}
\end{figure}

Interestingly, we see two peaks in the histogram, one between between $0\textrm{\AA}^3$ and $15\textrm{\AA}^3$, the second one above $15\textrm{\AA}^3$.~\cref{fig:render_lf} shows the atoms of the cross-linked network whose Voronoi regions have volume less than $15\textrm{\AA}^3$~(colored yellow) and the atoms whose Voronoi regions have volume greater than $15\textrm{\AA}^3$~(colored green). This image is of the sample containing pure DGEBA. Upon close observation, it is revealed that the atoms which are bonded to more than two atoms~(hydrogens excluded) fall into the category of atoms having volume lower than $15\textrm{\AA}^3$ and the atoms with two or less bonds fall in the category of atoms having volume higher than $15\textrm{\AA}^3$. 

In order to quantify the effect of the addition of BER, the mean of the Voronoi cell volume distribution has been computed.~\cref{fig:avmmean} shows the average Voronoi region volume as function of mass fraction of BER, as well as the individual averages for the Voronoi cells with volume below and above $15\textrm{\AA}^3$. It is seen that the addition of BER increases the Voronoi cell volume available to individual atoms. Furthermore, it is seen that the atoms whose Voronoi cells occupy greater than $15\textrm{\AA}^3$ are affected more drastically, as the slope of the linear fit for these atoms is approximately 34\% higher than for those atoms whose Voronoi cells occupy less than $15\textrm{\AA}^3$. 

\subsection{Shortest path between nitrogens}

\begin{figure}
     \centering
     \begin{subfigure}[b]{0.45\textwidth}
         \centering
         \includegraphics[width=\textwidth]{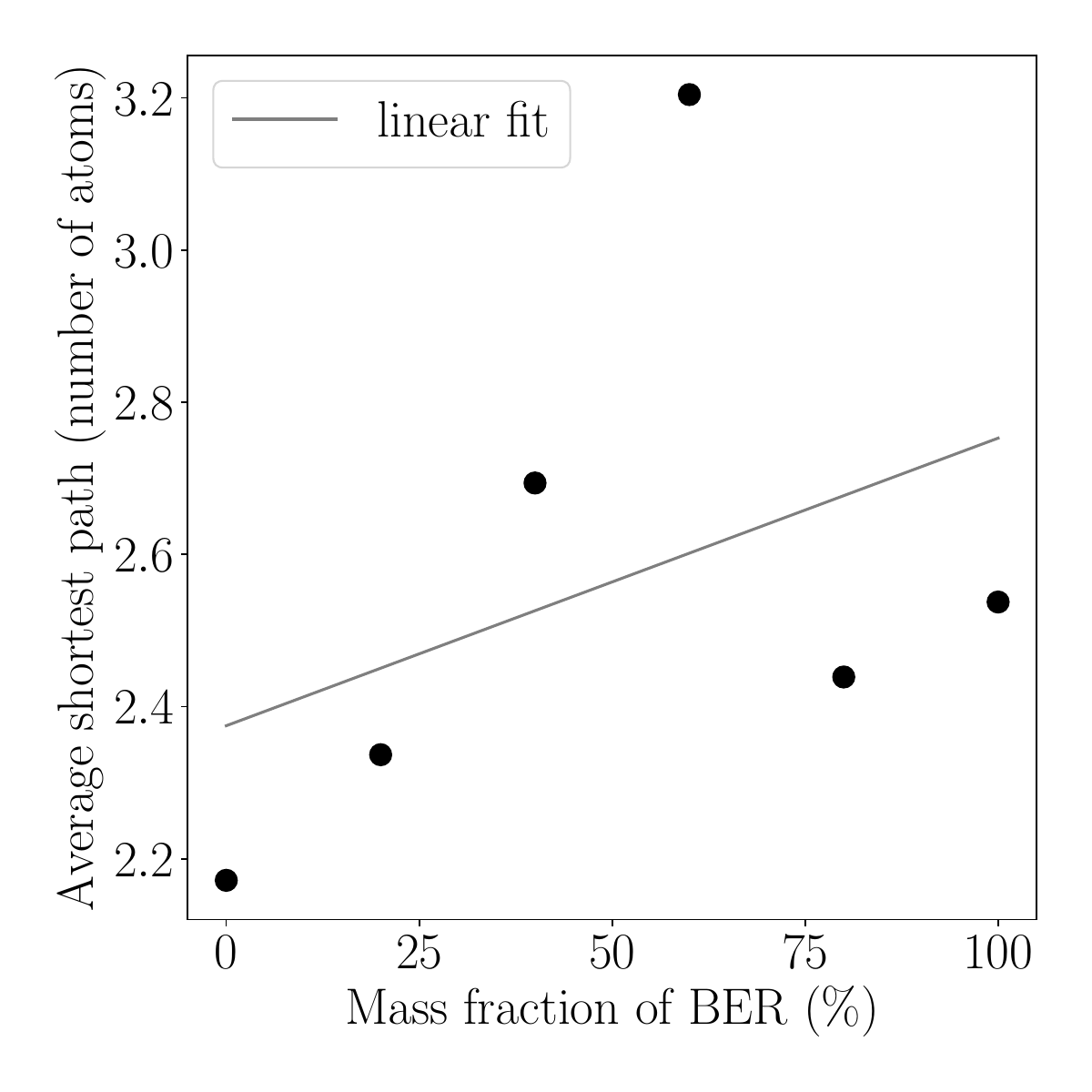}
         \caption{}
         \label{fig:sp_n}
     \end{subfigure}
     \begin{subfigure}[b]{0.45\textwidth}
         \centering
         \includegraphics[width=\textwidth]{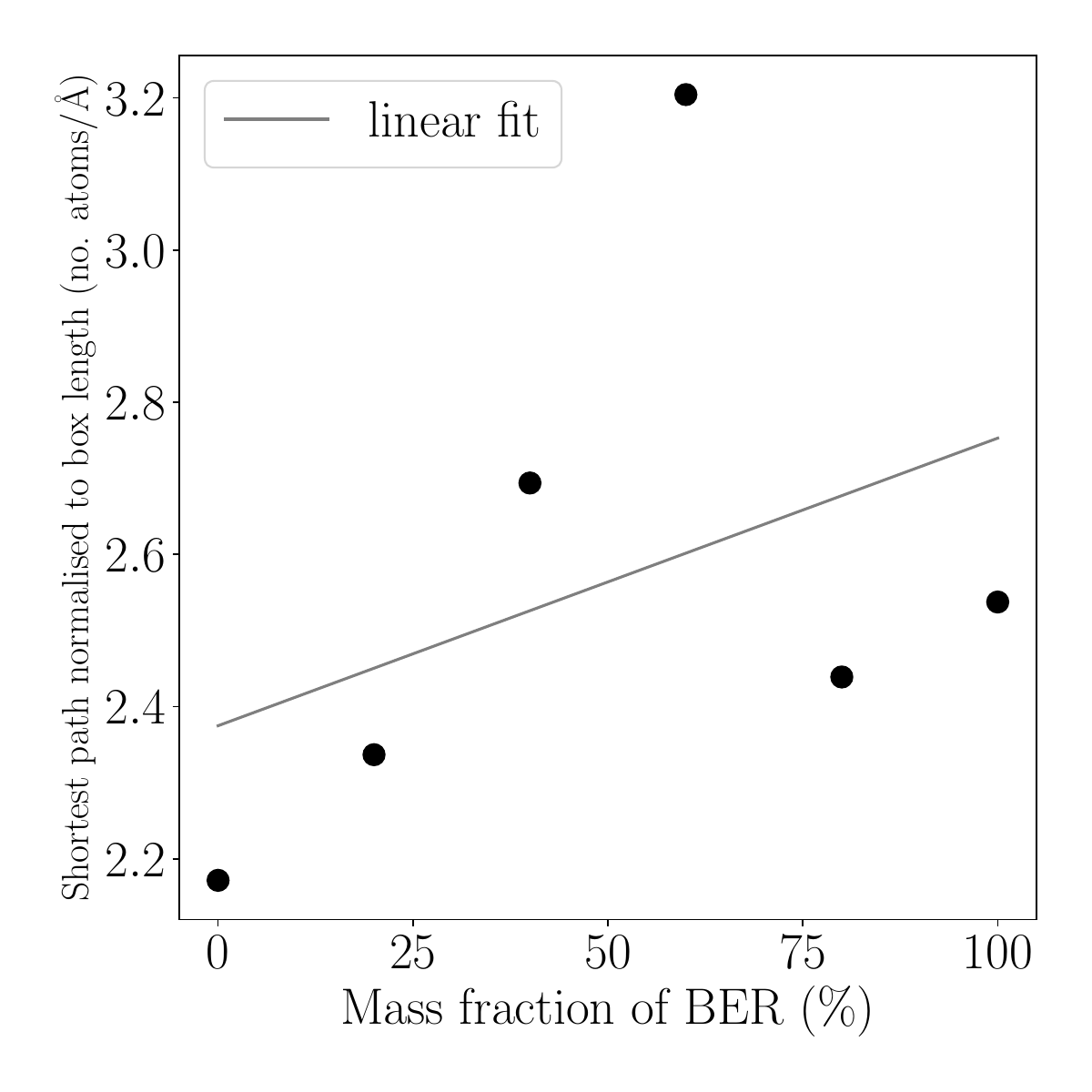}
         \caption{}
         \label{fig:sp_n_nml}
     \end{subfigure}
        \caption{(a) Average shortest path between nitrogens in the network versus the mass fraction of BER. (b) Average shortest path between nitrogens in the network normalized to the length of the simulation box versus the mass fraction of BER.}
        \label{fig:sp}
\end{figure}

We now look at the shortest path between the nitrogen atoms in the network.  Here, the shortest path between two atoms is defined as the least number of atoms that need to be traversed in order to reach one atom from other. The Dijkstra's algorithm~(\cite{dijkstra1959note}) has been utilized to find the shortest path between the Nitrogen atoms in the network. Nitrogen atoms were chosen as they are the sites where crosslinking occurs. If the network is visualized as a fishing net then the nitrogen atoms are akin to the knots in the net.~\cref{fig:sp_n} shows the average shortest path between the nitrogen atoms as a function of the mass fraction of BER. We see a positive correlation between the the mass fraction of BER and the average shortest length. Still, some part of this correlation is due to the increase in the size of the periodic box. Going back to the fishing net analogy, ~\cref{fig:sp_n} suggests that there is more thread between each of the knots. If the increase in the overall size of the net is similar to the increase in the amount of thread between knots, there is no change in the topology of the thread except for scale. This is why we also normalize the amount of thread between knots to the size of the net.~\cref{fig:sp_n_nml} shows the normalized shortest path. There is still positive correlation between the mass fraction of BER and the normalized average shortest path. This suggests that, in broad terms, the network is tightly packed with pure DGEBA, and as the amount of BER increases the network becomes looser.

\subsection{Confinement Index: A Measure of Atomic Packing}

In this subsection, we introduce a new metric to quantify how tightly packed a molecular system is, which we call the confinement index. We computed this confinement index by first assigning a freedom score to each atom in the network based on its bonding and local environment, as described below:

\begin{itemize}
    \item A free atom has a freedom score of 3.
    \item An atom bonded to one other atom has a freedom score of 2.
    \item An atom bonded to two atoms has a freedom score of 1.
    \item An atom bonded to three or more atoms has a freedom score 0.
    \item If an atom with freedom score 2 is bonded to an atom with freedom score 1, then its freedom score is reduced to 1.66.
    \item If an atom with freedom score 2 is bonded to an atom with freedom score 0, then its freedom score is reduced to 1.33.
    \item If an atom with freedom score 1 is bonded to one atom with freedom score 1 and one atom with freedom score 0 then its freedom score is reduced to 0.66.
    \item If an atom with freedom score 1 is bonded to two atoms with freedom score 0, then its freedom score is reduced to 0.33.
    \item Finally, the freedom score is divided by 1 + $N_c$ where $N_c$ is the number of atoms within a cutoff distance of the atom.
\end{itemize}

Once every atom is assigned the freedom score, the average freedom score of all the atoms is computed to give the average freedom score. We define confinement index as the inverse of the average freedom score.
Essentially, we are penalizing the presence of bonds, as well as crowding in the neighborhood while calculating the freedom score. If all the atoms were free and far apart, the average freedom score would be 3 and the confinement index would be $\frac{1}{3}$. Materials like diamond, in which every atom is bonded to four atoms each, would have an average freedom index of 0 and have infinite confinement index. Hence, confinement index can range from $\frac{1}{3}$ to infinity. The higher the confinement index, the more constrained the system is. One more important factor is to ensure we use the same cutoff distance while calculating the confinement index while comparing different systems. Too high cutoff distance would artificially drag down the freedom score, where as too little cutoff score would not capture the neighborhood crowding correctly. For the systems in this study, a cutoff distance of $3\textrm{\AA}$ was used and has been chosen after trial and error.

\begin{figure}
  \includegraphics[scale = 0.45]{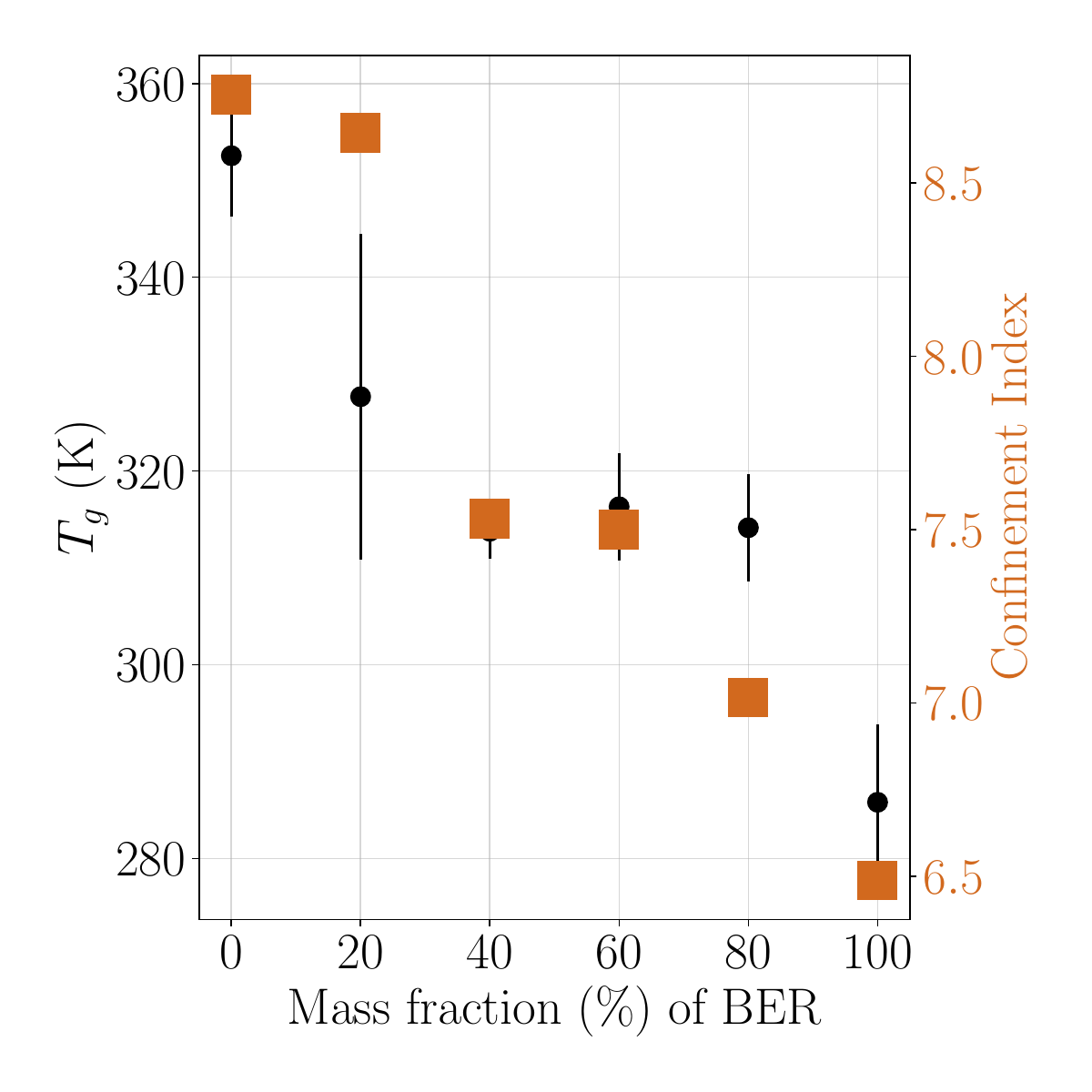}
  \caption{The glass-transition temperature along with the confinement index plotted against the mass fraction of BER.}
  \label{fig:citg}
\end{figure}

\cref{fig:citg} shows the glass-transition temperature of the samples and the  confinement index against the mass fraction of BER in the samples. A similar trend can be seen in both the properties. This suggest that the confinement index can probably be used as a heuristic for the glass-transition temperature. 

\subsection{Discussion}

Since topological effects are the most fundamental, they provide a natural starting point for the analysis. An increase in the inter-nitrogen path length with the addition of BER suggests that the network becomes progressively looser. This may result either from the bulky side chains of BER hindering dense packing or from the slightly longer backbone of BER compared to DGEBA. In either case, a looser network would lead to an increase in Voronoi volume, as reduced constraints allow the network to explore a wider range of conformations and adopt configurations that minimize steric overlap and potential energy.

The confinement index, which accounts for both topological looseness and local crowding, consequently decreases with increasing BER content. This reduction reflects the systematic decline in both its contributing factors: increased inter-nitrogen distances (network loosening) and enlarged Voronoi cell volumes (lesser neighborhood crowding).

An examination into the pattern at which RMSD levels off suggests that, above $T_g$, atomic mobility is mainly governed by molecular level rearrangement. The loosening of the network as well as the increase in atomic volume due to addition of BER has large impact on RMSD in the melt state and the RMSD saturation value increases dramatically with the increase in mass fraction of BER. However, below $T_g$, we observe that atomic movement is mainly governed by bond vibrations as RMSD levels off quickly and at a very small value. As topological loosening of the network has minimal effect on the bond scale atomic vibrations, there is minimal effect of the addition of BER in the solid state.

An important connection exists between RMSD and the coefficient of thermal expansion (CTE), as both are fundamentally driven by atomic mobility. Below $T_g$, thermal expansion is primarily governed by bond vibrations as is the case with RMSD. In this regime, the enhanced flexibility introduced by BER has limited impact. However, above $T_g$, thermal expansion is dominated by conformational changes within the network. Here, the increased flexibility of the looser network plays a more significant role, allowing the system to explore a broader range of configurations. As a result, the increase in CTE with increasing BER content is much more pronounced in the molten state.

The observed reduction in glass-transition temperature ($T_g$) is likely the result of multiple contributing factors. A key contributor is the looser network topology, which facilitates segmental mobility at lower temperatures. This occurs both directly, through an increased number of accessible conformations, and indirectly, by reducing atomic overlaps and effectively increasing the free volume available for motion.

On the other hand, the BER molecules introduce long side chains which, in principle, could increase $T_g$ by hindering the free rotation of the segments they are attached to. This effect would partially counteract the increased chain flexibility introduced by the looser network. Therefore, while it is difficult to attribute the reduction in $T_g$ to a single dominant factor, the overall trend emerges from the complex interplay between increased flexibility, free volume, and local steric hindrance. 

In calculating the confinement index, our aim was to develop a metric that captures certain local factors—such as steric hindrance and neighborhood crowding—which are known to influence molecular mobility. The resulting index serves as a rough qualitative heuristic for estimating trends in glass-transition temperature. Although the confinement index does not directly account for long-range chain flexibility or network-scale topological effects, it appears, at first glance, to correlate well with the observed trend in $T_g$. This suggests that even a locally derived parameter may hold predictive value in complex systems like crosslinked polymer networks.

\section{Conclusion}
In this study, we used molecular-dynamics simulations to examine how varying the resin composition—specifically the mass fraction of butylated epoxy resin (BER) in a DGEBA/DETDA system—affects the structural and thermal properties of crosslinked epoxy thermosets. A series of six systems with increasing BER content were simulated using a multi-step crosslinking approach, followed by thermal and structural characterization. 

Our results indicate that increasing the BER content leads to a measurable loosening of the network. This is supported by longer average nitrogen-nitrogen path lengths, increased Voronoi atomic volumes, and a reduction in the confinement index—a metric we propose to describe atomic-level crowding and connectivity. Together, these findings suggest that BER incorporation reduces local packing efficiency and network tightness.
 
Thermal analyses reveal that the glass-transition temperature ($T_g$) decreases with increasing BER content, while the coefficient of thermal expansion (CTE) increases. The increase in CTE is particularly pronounced in the melt state. Atomic mobility, as assessed through root mean square displacement (RMSD), also shows a clear increase above $T_g$ with BER content, consistent with a more flexible network. Below $T_g$, both CTE and RMSD show smaller changes, suggesting that bond-level vibrations dominate atomic motion in this regime, which are minimally influenced by network topology. 

While side groups in BER could potentially hinder local motion causing a reduction in $T_g$, the dominant effect influencing $T_g$ appears to be network loosening, leading to increased free volume and segmental mobility. The confinement index shows a trend similar to that of $T_g$, indicating its potential utility in capturing some aspects of local steric and topological effects. However, further studies would be required to evaluate its broader applicability.

Overall, this work highlights how the addition of BER affects network connectivity, local packing, and thermal response in epoxy thermosets. These findings contribute to the understanding of how resin composition can influence key properties of crosslinked polymers and may assist in guiding material design decisions where specific thermal or structural characteristics are desired.

\begin{acknowledgments}

This research was supported by the Netherlands Organization for Scientific Research (NWO) through funding provided under the REGENERATE~(KICH1.ST02.21.004) project.

The authors gratefully acknowledge Scienomics LLC for their guidance and support in working with the MAPS platform, which was instrumental in conducting the molecular-dynamics simulations presented in this study.

We also extend our sincere thanks to our partners in the REGENERATE project, especially Dr. Ir. J.A.W. (Hans) van Dommelen and Kylian A. van Akkerveken~(PhD Candidate) of the Department of Mechanical Engineering at Eindhoven University of Technology, as well as our industrial collaborators at ASML and Technotion, for their support, technical insights, and valuable guidance.

\end{acknowledgments}

\bibliography{main_references}

\end{document}